\documentclass[12pt]{article}
\usepackage{amssymb,slashed,latexsym,amsmath,multirow,color}
\usepackage{slashed}
\pdfoutput=1
\usepackage[font=footnotesize,labelsep=newline,labelfont=sc,justification=centering,position=top]{caption}

\numberwithin{equation}{section}

\def\be{\begin{equation}}
\def\ee{\end{equation}}
\def\bea{\begin{eqnarray}}
\def\eea{\end{eqnarray}}
\def\nn{\nonumber}
\def\tR{{\widetilde R}}
\def\tn{{\widetilde\nabla}}

\newcommand{\Symp}{\mathop{\rm {}Sp}}

\def\del{\partial}
\newcommand{\w}[1]{\\[0.#1cm]}
\newcommand{\eq}[1]{(\ref{#1})}

\newcommand{\opsi}{\bar{\psi}}

\usepackage{hyperref}

\newcommand{\ft}[2]{{\textstyle\frac{#1}{#2}}}
\def\rmi{{\rm i}}
\newsavebox{\uuunit}
\sbox{\uuunit}
    {\setlength{\unitlength}{0.825em}
     \begin{picture}(0.6,0.7)
        \thinlines
        \put(0,0){\line(1,0){0.5}}
        \put(0.15,0){\line(0,1){0.7}}
        \put(0.35,0){\line(0,1){0.8}}
       \multiput(0.3,0.8)(-0.04,-0.02){12}{\rule{0.5pt}{0.5pt}}
     \end {picture}}

\newcommand{\SU}{\mathop{\rm SU}}
\newcommand{\U}{\mathop{\rm U}}

\begin{document}
\begin{titlepage}
\begin{flushright}
RUG-12-23\\
MIFPA-11-56 \\
\vskip .1truecm

arXiv:1203.2975 [hep-th]
\end{flushright}
\vspace{.5cm}
\begin{center}
\baselineskip=16pt {\LARGE  Higher Derivative Extension of \w2 $6D$
Chiral Gauged Supergravity}
 \vskip 15mm 
{\large Eric Bergshoeff$^1$, Frederik Coomans$^2$,
Ergin Sezgin$^3$, Antoine Van Proeyen$^2$} \vskip .4truecm

\centerline{{\small  $^1$ Centre for Theoretical Physics, University of
Groningen, Nijenborgh 4,}}
 \centerline{{\small 9747 AG
Groningen, The Netherlands}}
\vskip .2truecm  \centerline{{\small $^2$Instituut voor Theoretische Fysica, Katholieke Universiteit
 Leuven,}}
      \centerline{{\small  Celestijnenlaan 200D B-3001 Leuven, Belgium}}
\vskip .2truecm
\centerline{{\small  $^3$ George and Cynthia Woods Mitchell
Institute for Fundamental Physics and Astronomy,}} \centerline{\small Texas
A\&M University, College Station, TX 77843, USA}  \vspace{6pt}

\end{center}
\vskip 1truecm

\begin{center}
{\bf Abstract}
\end{center}

Six-dimensional $(1,0)$ supersymmetric gauged Einstein--Maxwell
supergravity is extended by the inclusion of a supersymmetric Riemann
tensor squared invariant. Both the original model as well as the Riemann
tensor squared invariant are formulated off-shell and consequently the
total action is off-shell invariant without modification of the
supersymmetry transformation rules. In this formulation, superconformal
techniques, in which the dilaton Weyl multiplet plays a crucial role,
are used.  It is found that the gauging of the $\U(1)$ R-symmetry in the
presence of the higher-order derivative terms does not modify the
positive exponential in the dilaton potential. Moreover, the
supersymmetric Minkowski$_4 \times S^2$ compactification of the original
model, without the higher-order derivatives, is remarkably left intact.
It is shown that the model also admits non-supersymmetric vacuum
solutions that are direct product spaces involving de Sitter spacetimes
and negative curvature internal spaces.

 \vfill

\hrule width 3.cm \vspace{2mm}{\footnotesize \noindent e-mails:
E.A.Bergshoeff@rug.nl, \{Frederik.Coomans, Antoine.VanProeyen\}@fys.kuleuven.be, \\ Sezgin@physics.tamu.edu}
\end{titlepage}
\addtocounter{page}{1}
 \tableofcontents{}
\newpage

\section{Introduction}
Higher-order curvature terms in supergravity theories are of
considerable importance for different reasons. They can be considered as
higher-order correction terms (in $\alpha'$) to an effective
supergravity Lagrangian of a (compactified) string theory (see, e.g.,
\cite{Bergshoeff:1989de}). These Lagrangians are supersymmetric only
order by order in the perturbation parameter $\alpha^\prime$. On the
other hand off-shell formulations for different curvature squared
invariants in 4, 5 and 6 dimensions have been constructed in
\cite{Bergshoeff:1986wc, Bergshoeff:1987rb, Argyres:2003tg,
Hanaki:2006pj, deWit:2010za, Bergshoeff:2011xn}. These invariants, added
to a pure off-shell supergravity theory, are \textit{exactly}
supersymmetric  and can be  considered in their own right. The off-shell
nature of these theories implies that they contain auxiliary fields. It
is well-known that, when adding higher derivative terms to the
Lagrangian, the auxiliary fields become propagating. Hence, the
elimination of these auxiliary fields becomes much harder since their
field equations are not algebraic anymore. Assuming that the
dimensionful parameter in front of the higher derivative part of the
Lagrangian is very small, one can solve the auxiliary field equations
{\sl perturbatively} and eliminate these fields order by order in the
small parameter. It remains an open question if and how the on-shell
Lagrangian obtained in this way is related to the compactified string
Lagrangian, which does not contain any auxiliary fields to begin
with.\footnote{The elimination of auxiliary fields in higher derivative
theories has been discussed in \cite{Argyres:2003tg}. A conjectured
duality between a supergravity Lagrangian with the auxiliary fields
eliminated perturbatively and a compactified string Lagrangian, without
auxiliary fields, can be found in section 5 of \cite{Lu:2010ct}.}

Theories containing higher-order curvature terms can provide corrections
to black hole entropies \cite{Mohaupt:2000mj, Cremonini:2008tw,
deWit:2009de} and can source higher-order effects in the AdS/CFT
correspondence \cite{Liu:2010gz, Cremonini:2009sy}. When considering
these theories as toy models on their own they can be compactified to
lower dimensions. A particular case to consider is the compactification
to three dimensions \cite{Lu:2010ct}. A particular feature of three
dimensions is that $D=3$ gravitons are non-propagating when only
considering 2-derivative Lagrangians. Instead, the addition of
higher-derivative terms can turn these non-propagating modes into
propagating {\sl massive} graviton modes, see e.g. \cite{Bergshoeff:2010mf} and references therein. These theories can then
be regarded as simple toy models to study quantum gravity.

In this paper we study higher-order corrections to a six-dimensional
$(1,0)$ supersymmetric $\U(1)_R$ gauged Einstein--Maxwell supergravity
theory, usually referred to as the Salam--Sezgin model
\cite{Salam:1984cj}, which is a special case of a $\Symp(n)\times
\Symp(1)_R$ gauged matter-coupled supergravity theory that was first
obtained in \cite{Nishino:1984gk}. We shall refer to this more general
case as  $6D$ chiral gauged supergravity as well. An intriguing feature
of the Salam--Sezgin model is that it allows a compactification over
$S^2$ to a four-dimensional Minkowski spacetime while retaining half of
the supersymmetry \cite{Salam:1984cj}. One of the purposes of this work
is to investigate whether this feature survives after the addition of
higher-order derivative corrections. To facilitate the addition of such
higher-order corrections to the model we will first construct its
off-shell formulation. It turns out that this is only possible for the
dual formulation of the model where the 2-form potential ${\tilde B}$
has been replaced by a dual 2-form potential $B$ \cite{Nishino:1997ff,Ferrara:1997gh}.
This has the effect that the curvature of the original 2-form potential
no longer contains a Maxwell--Chern--Simons term, but that instead a
term of the form $ B \wedge F \wedge F$, where $F$ is the Maxwell field
strength, appears in the Lagrangian.

To construct the off-shell formulation we will make use of the
superconformal tensor calculus. As a first step we will  review the
construction of off-shell minimal $D=6$ supergravity
\cite{Bergshoeff:1985mz, Coomans:2011ih}. In this construction one makes
use of the dilaton Weyl multiplet (obtained by coupling the regular Weyl
multiplet to a tensor multiplet) coupled to a linear multiplet as compensator. After
fixing the conformal symmetries, this theory still has a remaining
$\U(1)$ R-symmetry which is gauged by an auxiliary vector
$\cal{V}_{\mu}$. We will couple this `pure' theory to an Abelian vector
multiplet and show that after solving for the auxiliary $\cal{V}_{\mu}$,
the gauging proceeds via the vector $W_{\mu}$ of the Abelian vector multiplet.

After constructing the off-shell formulation of the gauged $(1,0)$
supergravity theory, we investigate its deformation by an off-shell curvature
squared invariant \cite{Bergshoeff:1986wc, Bergshoeff:1987rb}.  To
construct this invariant it is essential to make use of the dilaton Weyl
multiplet. We review the construction of this higher-derivative term and
add it to the off-shell $(1,0)$ supergravity theory. Next, we study the gauging
procedure in the presence of the Riemann tensor squared invariant.

As a first step towards understanding the properties of the higher-derivative extension of the model we perform a systematic search for
vacuum solutions. We construct both supersymmetric as well
as non-supersymmetric solutions. For one particular supersymmetric
solution, namely six-dimensional Minkowski spacetime, we calculate the
fluctuations around this background and show how these fluctuations fit into
supermultiplets.

This paper is organized as follows. In section \ref{section: R-symmetry
gauging} we review the off-shell version of the $(1,0)$ supergravity
model
\cite{Bergshoeff:1985mz,Coomans:2011ih} and describe its gauging. In
Section \ref{app:totalL}, we introduce an alternative off-shell
formulation of the model in view of the fact that it is best suited for
the addition of the Riemann tensor squared invariant \cite{Bergshoeff:1986wc}.
In section \ref{Section:gaugingwithR2terms} we discuss the construction
of the Riemann tensor squared invariant and arrive at the total Lagrangian for
the higher-derivative extended $6D$ chiral gauged supergravity theory. In Section
\ref{ss:vacuumsol}, we investigate the vacuum solutions of this model.
We summarize and comment further on our results and on some interesting open
problems in the Conclusions section. Throughout the paper we follow the
notation given in Appendix A of \cite{Coomans:2011ih}.

\section{Off-shell Gauged \texorpdfstring{$(1,0)$}{(1,0)} Supergravity} \label{section: R-symmetry gauging}

In this section we present an off-shell version of the dual formulation
\cite{Nishino:1997ff,Ferrara:1997gh} of the Salam--Sezgin model
\cite{Salam:1984cj,Nishino:1984gk}. In the first subsection we give the
off-shell Lagrangian of pure supergravity plus a tensor multiplet as
constructed in \cite{Bergshoeff:1985mz, Coomans:2011ih}. In the next
subsection we couple a vector multiplet to this theory and show that the
resulting Einstein--Maxwell model leads to a non-trivial $\U(1)$ gauge
symmetry that is not gauged by an auxiliary vector field. In the last
subsection we show that after eliminating the auxiliary fields one ends
up with a Lagrangian in which the $\U(1)$ gauge symmetry is effectively
gauged by the physical vector of the vector multiplet. We furthermore
show that, after dualizing the 2-form potential into a dual 2-form
potential, this Einstein--Maxwell model is nothing else than the original
Salam--Sezgin model.


\subsection{Off-shell Poincar{\'e} Action} \label{ss:Poincaction}


The off-shell $(1,0)$ supergravity action has been constructed by means of a superconformal
tensor calculus in which the off-shell so-called dilaton Weyl multiplet with independent fields
\begin{equation}
\{\,e_{\mu}{}^a\ ,\psi_\mu^i, B_{\mu\nu}\ , {\cal V}_\mu^{ij}\ ,b_\mu\ ,\psi^i\ , \sigma\, \}
\label{Weyl}
\end{equation}
and Weyl weights $(-1,-1/2, 0,0,0,5/2,2)$, respectively, is coupled to an off-shell linear
multiplet consisting of the fields
\be
\{\,E_{\mu\nu\rho\sigma}\ , L^{ij}\ , \varphi^i\, \}\ ,
\label{linear}
\ee
with Weyl weights $(0,4,9/2)$, respectively. The fields $(\psi_\mu^i,
\psi^i, \varphi^i)$ are symplectic Majorana--Weyl spinors labelled by a
$\Symp(1)_R$ doublet index, the fields $B$ and $E$ are two- and
four-forms with tensor gauge symmetries, respectively, $b_\mu$ is the
dilatation gauge field and $L_{ij}$  are three real scalars. An
appropriate set of gauge choices for obtaining  off-shell supergravity
with the Einstein--Hilbert term, namely ${\mathcal L}= eR+\cdots$, is
given by
\be
\boxed{
L_{ij} =\frac{1}{\sqrt 2}\,\delta_{ij}\ , \quad \varphi^i=0\ , \quad b_\mu=0}
\label{gf1}
\ee
which fixes the dilatations, conformal boost and special supersymmetry transformations. Moreover,
the first of the gauge choices in \eq{gf1} breaks $\Symp(1)_R$  down to $\U(1)_R$.
This set of gauge choices leads to an off-shell multiplet containing $48+48$  degrees of freedom described by the fields \cite{Bergshoeff:1985mz}
(see Table~5 of \cite{Coomans:2011ih})
\begin{equation}e_{\mu}{}^a\ (15)\,,\ \ {\cal V}^\prime_{\mu}{}^{ij}\
(12)\,,\ \ {\cal V}_\mu\ (5)\,,\ \ B_{\mu\nu}\
(10)\,,\ \ \sigma\ (1)\,,\ \ E_{\mu\nu\rho\sigma}\ (5)\,;\ \ \psi_{\mu}{}^i\ (40)\,,\ \ \psi^i \
(8)\,.\label{m1}
\end{equation}
 The field ${\cal V}_\mu$ is the gauge field of the surviving $\U(1)_R$ gauge symmetry. It arises
in the decomposition
\begin{equation}
{\cal V}_\mu^{ij}={\cal V}_\mu^{\prime ij}+\frac12\delta^{ij} {\cal V}_\mu\ ,\qquad {\cal V}_\mu^{\prime ij}\,\delta_{ij}=0\ ,
\label{decomposition}
 \end{equation}
where the traceless part ${\cal V}_\mu^{\prime ij}$ has no gauge symmetry.
A superconformal tensor calculus method was employed in \cite{Bergshoeff:1985mz} where the bosonic action was given, and a procedure for obtaining the full action was provided. This full action, including the quartic terms, was constructed in  \cite{Coomans:2011ih}.
The Lagrangian up to quartic fermion terms is given by \cite{Bergshoeff:1985mz,Coomans:2011ih}
\footnote{We use the conventions of \cite{Coomans:2011ih}. In particular, the spacetime signature is $(-+++++)$,
$\gamma_{a_1\cdots a_6}= \varepsilon_{a_1\cdots a_6}\gamma_*$, $\gamma_*\epsilon=\epsilon$, ${\bar\psi}_i \psi_j= -{\bar\psi}_j\psi_i$ and ${\bar\psi}_i \gamma_\mu\psi_j= {\bar\psi}_j \gamma_\mu\psi_i$.
These conventions differ from those in \cite{Bergshoeff:1985mz} in using signature $(-+\ldots +)$ rather than the
Pauli convention $(++\ldots +)$,
in rescaling ${\cal V}_\mu^i{}_j$ by a factor of $-1/2$, and the minus sign in the definition of the Ricci tensor. The signature change merely results in rescaling $\varepsilon^{\mu_1...\mu_6}$ by a factor of $\rmi$.
}
\begin{eqnarray}
\left.e^{-1}{\mathcal L}_R\right|_{L=1}
&=& \frac12 R-\frac12\sigma^{-2}\partial_\mu\sigma\partial^\mu\sigma
-\frac{1}{24}\sigma^{-2}F_{\mu\nu\rho}(B) F^{\mu\nu\rho}(B)+{\cal V}'_{\mu ij}{\cal V}^{\prime \mu ij} \nonumber\\
 &&-\frac14 E^\mu E_\mu +\frac{1}{\sqrt 2} E^\mu {\cal V}_\mu -\frac{1}{4\sqrt 2} E_\rho \opsi_\mu^i \gamma^{\rho\mu\nu} \psi_\nu^j \delta_{ij}
 \nonumber \\
&&- \frac{1}{2}\opsi_\mu\gamma^{\mu\nu\rho}D_\nu(\omega )\psi_\rho
-2\sigma^{-2}{\bar\psi}\gamma^\mu D'_\mu(\omega )\psi+\sigma^{-2}\opsi_\nu\gamma^\mu\gamma^\nu \psi\,\partial_\mu \sigma
\label{offshellPoincare} \\
&& -\frac{1}{48} \sigma^{-1} F_{\mu\nu\rho}(B) \left( \opsi^\lambda\gamma_{[\lambda} \gamma^{\mu\nu\rho}\gamma_{\tau]}\psi^\tau
+4 \sigma^{-1} \opsi_\lambda\gamma^{\mu\nu\rho}\gamma^\lambda\psi -4\sigma^{-2} {\bar\psi}\gamma^{\mu\nu\rho}\psi\right)\ .
\nonumber
\end{eqnarray}
The indication $L=1$ in the left-hand side indicates all the gauge choices
(\ref{gf1}).
Here we have defined the field strength for the 2-form potential and the dual of the field strength for the
 4-form potentials as follows\footnote{Note that the definition of $E^\mu$ here is purely bosonic, and it differs from the definition used in
 \cite{Bergshoeff:1985mz,Coomans:2011ih}, where it is a superconformal covariant
 expression with fermionic bilinear terms.}
\bea
F_{\mu\nu\rho}(B) &=& 3\partial_{[\mu} B_{\nu\rho]}\ ,
\label{fb}\w2
\qquad E^\mu &=& \frac{1}{24}e^{-1}\varepsilon^{\mu\nu_1\cdots\nu_5} \partial_{[\nu_1}E_{\nu_2\cdots \nu_5]}\ .
\label{defFBE}
\eea
The $\U(1)_R$
covariant derivatives $D_\mu(\omega ) $ and the full $\SU(2)$ covariant derivatives $D'_\mu(\omega )$
are given by
\bea
D_\mu(\omega )\psi_\nu^i &=& (\partial_\mu + \frac14\omega_{\mu}{}^{ab}\gamma_{ab} )\psi_\nu^i
-\frac12 {\cal V}_\mu\delta^{ij}\psi_{\nu j}\ ,
\label{cd1}\w2
D'_\mu(\omega )\psi^i &=& (\partial_\mu + \frac14\omega_{\mu}{}^{ab}\gamma_{ab})\psi^i
-\frac12 {\cal V}_\mu\delta^{ij}\psi_j + {\cal V}_\mu{}^{\prime i}{}_j \psi^j\ ,
\label{cd2}
\eea
where $\omega_{\mu ab}$ is the standard torsion-free connection. Note that the symmetric traceless field ${\cal V}'_{\mu}{}^{ij}$, occurring in the decomposition
(\ref{decomposition}), is absent in the covariant derivative of the gravitino \cite{Coomans:2011ih}.
This is a consequence of having broken the $\SU(2)$ symmetry present in the dilaton Weyl multiplet by the gauge choices \eq{gf1}.
In the above formula, and throughout the paper the spin connection $\omega_{\mu ab}$ is the standard one associated with the
Christoffel symbol, and as such, it does not depend on fermionic or bosonic torsion. The supersymmetry transformations,
up to cubic fermion terms, are obtained from Sec. 2 of
\cite{Coomans:2011ih}:
\begin{eqnarray}
\delta e_{\mu}{}^a &=& \frac{1}{2}\bar{\epsilon}\gamma^a\psi_{\mu}\,,
\nonumber \\
\delta \psi_{\mu}^i &=& D_\mu(\omega )\epsilon^i +\frac{1}{48}\sigma^{-1}\gamma\cdot F(B) \gamma_\mu \epsilon^i
-{\cal V}_\mu^{\prime ij}\epsilon_j +\gamma_{\mu}\eta^i\ ,
\nonumber \\
\delta B_{\mu\nu} &=& -\sigma \bar{\epsilon}\gamma_{[\mu}\psi_{\nu]}-\bar{\epsilon}\gamma_{\mu\nu}\psi\,,
\nn\w2
\delta \psi^i &=& \frac{1}{48}\gamma \cdot F(B)\epsilon^i+\frac{1}{4}\slashed{\partial}\sigma\epsilon^i-\sigma\eta^i\,,
\nn\w2
\delta \sigma &=& \bar{\epsilon}\psi\,,
\label{susy1}\w2
\delta E_{\mu\nu\rho\sigma} &=& 2\sqrt{2}\opsi_{[\mu}{}^i\gamma_{\nu\rho\sigma]}\epsilon^j\delta_{ij}\,,
\nn\w2
\delta {\cal V}_{\mu}^{ij} &=& \frac{1}{2}\bar{\epsilon}^{(i}\gamma^\nu R_{\mu\nu}{}^{j)}(Q)
+\frac{1}{8}\sigma^{-1}\bar{\epsilon}^{(i}\gamma^\nu \Bigl({F}_{[\mu}{}^{ab}(B)\gamma_{ab}\psi_{\nu] }^{j)}\Bigr)+\frac{1}{24}\sigma^{-1}\bar{\epsilon}^{(i}\gamma\cdot F(B)\psi_{\mu}^{j)}
\nonumber \\
&&
+\frac{1}{2}\sigma^{-1}\bar{\epsilon}^{(i}\gamma_{\mu}\slashed{D}'(\omega )\psi^{j)}-\frac{1}{8}\sigma^{-1}\bar{\epsilon}^{(i}\gamma_{\mu}\gamma^{\rho}\slashed{\partial}\sigma\psi_{\rho}{}^{j)}-\frac{1}{48}\sigma^{-2}\bar{\epsilon}^{(i}\gamma_{\mu}\gamma\cdot
F(B)\psi^{j)}+2\bar{\eta}^{(i}\psi_{\mu}^{j)}\,, \nonumber
\end{eqnarray}
where $D_\mu(\omega )\epsilon^i$ is defined as in \eq{cd1}, $R_{\mu\nu}{}^i(Q)$ is the gravitino
curvature and $\eta _i$ is the effective contribution from the
$S$-supersymmetry in the superconformal algebra:
\begin{eqnarray}
 D_\mu(\omega )\epsilon^i & = & (\partial_\mu + \frac14\omega_{\mu}{}^{ab}\gamma_{ab} )\epsilon ^i
-\frac12 {\cal V}_\mu\delta^{ij}\epsilon _j\,, \nonumber\\
 R_{\mu\nu}{}^i(Q) & = &2D_{[\mu }(\omega )\psi _{\nu  ]}^i -2{\cal V}_{[\mu }^{\prime\,ij}\psi _{\nu  ]j}\,,\label{rq}\w2
\eta_k &=&\frac14\Bigl( \gamma^{\mu}{\cal V}_{\mu}{}^{\prime (i}{}_l\delta^{j)l}\epsilon_j
-\frac{1}{2\sqrt 2} E_\mu \gamma^\mu \epsilon^i\Bigr)\delta_{ik}\ .
\label{eta}
\end{eqnarray}
The latter equation gives the compensating special supersymmetry transformation parameter
in the gauge $\varphi^i=0$, as can be read off from eq. (3.37) of
\cite{Bergshoeff:1985mz}. Note that the $\U(1)_R$ part of ${\cal
V}_\mu^{ij}$ has dropped out in this expression. The surviving $\U(1)_R$
symmetry of the Lagrangian $\mathcal{L}_R$ is gauged by the auxiliary
gauge field ${\cal V}_\mu$, which acts as follows \,\footnote{The
$\U(1)_R$ is the subgroup of the full $\rm{SU}(2)$, under which the
gravitino transforms as
$$\delta
\psi_{\mu}{}^i=-\lambda^i{}_j\psi_{\mu}{}^j=\bigl(\lambda'^{ij}+\frac{1}{2}\lambda\delta^{ij}\bigr)\psi_{\mu
j}\,,$$ where $\lambda'^{ij}$ is traceless. A similar formula holds for
$\psi^i$.}
\begin{equation}
\delta(\lambda) {\cal V}_{\mu}=\partial_{\mu}\lambda\,,\qquad
\delta(\lambda) \psi_{\mu}{}^i=\frac{1}{2}\delta^{ij}\lambda\psi_{\mu j}\,,\qquad
\delta(\lambda) \psi^i=\frac{1}{2}\delta^{ij}\lambda\psi_j\,,
\end{equation}
with $\lambda$ being the parameter of the gauged symmetry.


\subsection{Coupling to an Off-shell Vector Multiplet} \label{sect:Rsymmgauging}

We now wish to introduce a gauge multiplet, whose vector is {\sl not}
auxiliary, to gauge the $\U(1)$ R-symmetry. The present gauging by
${\cal V}_{\mu}$, discussed in the previous subsection, is undesirable
since ${\cal V}_{\mu}$ has no standard kinetic term. In fact, we will
show in subsection \ref{ss:elimAuxF} that the gauge symmetry becomes
trivial after solving the 4-form potential in terms of a scalar field.

To obtain this non-trivial gauging we follow \cite{Bergshoeff:1985mz}
and add to $\mathcal{L}_{\rm{R}}$ the kinetic terms for an abelian
vector multiplet $\mathcal{L}_{\rm{V}}$. The multiplet consists
of the fields $(W_\mu, Y_{ij}, \Omega _i)$, being a physical gauge
field, an auxiliary $\SU(2)$ triplet, and a physical fermion. They
transform under dilatations with Weyl weights $(0,2,3/2)$, respectively.
We add the coupling $g\mathcal{L}_{\rm{VL}}$ of the vector
multiplet to the compensating linear multiplet. Prior to fixing any of
the conformal symmetries, these Lagrangians, up to quartic fermion
terms, are given by \cite{Bergshoeff:1985mz}
\begin{eqnarray}
e^{-1}\mathcal{L}_{\rm{V}}&=&\sigma\Bigl(-\frac{1}{4}F_{\mu\nu}(W)F^{\mu\nu}(W)
-2\bar{\Omega}\gamma^\mu D'_\mu(\omega ) \Omega +Y^{ij}Y_{ij}\Bigr) \nonumber \\
&&-\frac{1}{16}e^{-1}\varepsilon^{\mu\nu\rho\sigma\lambda\tau}B_{\mu\nu}F_{\rho\sigma}(W)F_{\lambda\tau}(W)
-4\bar{\Omega}^i\psi^jY_{ij} \nonumber \\
&&+\frac12 \left(\sigma \bar{\Omega }\gamma^{\mu}\gamma\cdot F(W) \psi_\mu
+2{\bar\Omega }\gamma\cdot F(W)\psi\right)+\frac{1}{12}\bar{\Omega }\gamma\cdot F(B)\Omega\ ,
\label{abelianvector}\w2
e^{-1}\mathcal{L}_{\rm{VL}} &=& Y_{ij}L^{ij}+2\bar{\Omega}\varphi
-L^{ij}\opsi_{\mu i}\gamma^{\mu}\Omega_j + \frac12 W_\mu E^\mu\,,
\label{LVLconf}
\end{eqnarray}
where $D'_\mu(\omega )\Omega ^i$ is defined as in \eq{cd2}. This action has the
full $\SU(2)$ symmetry.

The coupling of the vector multiplet to supergravity is then achieved by considering
the Lagrangian
\be
{\mathcal L}_1= \Bigl( {\mathcal L}_R + {\mathcal L}_V
+g{\mathcal L}_{VL} \Bigr)\Big\vert_{L=1}\ ,
\ee
where as before `$L=1$' refers to the set of gauges given in \eq{gf1}. This formula, up to quartic fermion terms,
yields the result
\begin{eqnarray}
e^{-1}{\mathcal L}_1 &=& \frac{1}{2}R -\frac{1}{2}\sigma^{-2}\partial_\mu \sigma\partial^\mu \sigma
+\frac{1}{\sqrt{2}}g \delta^{ij}Y_{ij} -\frac{1}{24}\sigma^{-2}F_{\mu\nu\rho}(B)F^{\mu\nu\rho}(B)
\nn\\
&&  + {\cal V}'_\mu{}^{ij}{\cal V}^{\prime \mu}{}_{ij} -\frac14 E^\mu E_\mu
+\frac{1}{\sqrt 2}E^\mu ({\cal V}_\mu+\frac{1}{\sqrt{2}}g W_\mu)
\nn\\
&& +\sigma Y^{ij} Y_{ij}-\frac{1}{4}\sigma F_{\mu\nu}(W)F^{\mu\nu}(W)
-\frac{1}{16}e^{-1}\varepsilon^{\mu\nu\rho\sigma\lambda\tau}B_{\mu\nu}F_{\rho\sigma}(W)F_{\lambda\tau}(W)
\nn\w2
&& -\frac{1}{2}\opsi_{\rho}\gamma^{\mu\nu\rho}D_{\mu}(\omega )\psi_{\nu}
-2\sigma^{-2}{\bar\psi}\gamma^\mu D'_\mu(\omega ) \psi
 +\sigma^{-2}\opsi_\nu\gamma^\mu\gamma^\nu\psi\partial_\mu\sigma
\nn\\
&&  -\frac{1}{48}\sigma^{-1} F_{\mu\nu\rho}(B) \Bigl(\opsi^\lambda\gamma_{[\lambda} \gamma^{\mu\nu\rho}\gamma_{\tau]}\psi^\tau
+4 \sigma^{-1} \opsi_\lambda\gamma^{\mu\nu\rho}\gamma^\lambda\psi
-4\sigma^{-2} {\bar\psi}\gamma^{\mu\nu\rho}\psi\Bigr)
\nn\\
&& -\frac{1}{4\sqrt 2} E_\rho \psi_\mu^i \gamma^{\rho\mu\nu} \psi_\nu^j \delta_{ij}
-\frac{1}{\sqrt 2}g \delta^{ij} {\bar\Omega}_i\gamma^\mu\psi_{\mu j}
-2\sigma{\bar\Omega}\gamma^\mu D'_\mu(\omega )\Omega -4 Y^{ij}{\bar\Omega}_i\psi_j
\nn\\
&& +\frac12F_{\mu\nu}(W) \left(\sigma {\bar\Omega}\gamma^\lambda\gamma^{\mu\nu}\psi_\lambda
+2{\bar\Omega}\gamma^{\mu\nu}\psi\right)
+\frac{1}{12}F_{\mu\nu\rho}(B){\bar\Omega}\gamma^{\mu\nu\rho} \Omega\ .
\label{Lag1}
\end{eqnarray}
The action corresponding to the Lagrangian ${\mathcal L}_1$ is invariant
under the supersymmetry transformations \eq{susy1} supplemented by the
supersymmetry transformations of the components of the off-shell vector
multiplet. The transformations of the latter are given up to
cubic fermion terms by \cite{Bergshoeff:1985mz}
\begin{eqnarray}
 \delta W_\mu &=& -{\bar\epsilon}\gamma_\mu\Omega\ ,
\nn\\
\delta \Omega^i &=& \frac18 \gamma\cdot F(W)\epsilon^i -\frac12
Y^{ij}\epsilon_j\ ,
\nn\\
\delta Y^{ij} &=& -\frac12 {\bar\epsilon}^i \gamma^\mu\left(D'_\mu(\omega ) \Omega^j
-\frac18 \gamma\cdot F(W)\psi_\mu^j+\frac12 Y^{jk}\psi_{\mu k}\right)
+{\bar\eta}^i\Omega^j + ( i\leftrightarrow j)\ ,
\label{susyYM1}
\end{eqnarray}
where $\eta$ is as defined in \eq{eta}.  The Lagrangian ${\mathcal L}_1$
also has a manifest $\U(1)_R\times \U(1)$ symmetry with transformations
parametrized by $\lambda$ and $\eta$
\begin{eqnarray}
\delta {\cal V}_{\mu}&=&\partial_{\mu}\lambda\,, \qquad \delta W_{\mu}=\partial_{\mu}\eta, \nn\w2
\delta\psi_\mu^i &=& \frac12
\lambda \delta^{ij} \psi_{\mu j}\ , \qquad \delta\psi^i= \frac12
\lambda \delta^{ij}\psi_{\mu j}\ , \qquad \delta\Omega^i= \frac12
\lambda \delta^{ij}\Omega_{j}\ , \label{doubletransformations}
\end{eqnarray}
where $(\lambda \,,\eta )$ are the parameters of the $\big(\U(1)_R\,,\U(1)\big)$ symmetry, respectively.


\subsection{Elimination of Auxiliary Fields} \label{ss:elimAuxF}


We consider Lagrangian ${\cal L}_1$ given in (\ref{Lag1}), and begin by
writing down the field equations for the auxiliary fields $Y_{ij}, {\cal
V}_\mu^{\prime ij}, {\cal V}_\mu, E_{\mu\nu\rho\sigma}$:
\bea 0&=& \sigma Y_{ij}+\frac{1}{2\sqrt 2} g\delta_{ij} -2\bar \Omega_{(i}\psi_{j)}\ ,
\label{aux1}\w2
0&=& {\cal V}_\mu^{\prime ij} + \left(\sigma^{-2}{\bar\psi}^i\gamma_\mu\psi^j
+\sigma{\bar\Omega}^i\gamma_\mu\Omega^j -{\rm trace} \right)\ ,
\label{aux2}\w2 0&=& E^\mu +{\sqrt 2}\delta ^{ij}\left(\frac{1}{4}\bar \psi _{\nu i }\gamma ^{\mu \nu \rho }\psi _{\rho j}
-\sigma^{-2}{\bar\psi}_i\gamma^\mu\psi_j
-\sigma{\bar\Omega}_i\gamma^\mu\Omega_j\right) \ , \label{aux3}\w2
0&=& \varepsilon^{\lambda\tau\rho\sigma\mu\nu}\partial_\mu \left(
E_\nu -{\sqrt 2}{\cal V}_\nu-g W_\nu +\frac{1}{2\sqrt 2} {\bar\psi}^{\alpha i} \gamma_{\nu\alpha\beta} \psi^{\beta j} \delta_{ij}
\right)\ .
\label{aux4}
\eea
The elimination of $Y_{ij}$ in \eq{Lag1} by means of \eq{aux1} gives a positive
definite potential $\frac14 g^2\sigma^{-1}$ and the elimination of
${\cal V}_\mu^{\prime ij}$ by means of \eq{aux2} gives only quartic fermion
terms in the action. Next, \eq{aux4} implies that locally we can write
\be
E_\mu -{\sqrt 2}{\cal V}_\mu-g W_\mu
+\frac{1}{2\sqrt 2} {\bar\psi}^{\nu  i} \gamma_{\mu\nu \rho } \psi^{\rho  j} \delta_{ij}
 =\partial_\mu \phi\ ,
\label{c1} \ee
for some scalar field $\phi$ transforming under the $\U(1)_R\times \U(1)$ transformations
\eq{doubletransformations}
as
\be
\delta\phi = -g\eta - \sqrt 2 \lambda \ .
\label{gt}
\ee
The terms in \eq{c1} can be rearranged to write
\be
E_\mu = D_\mu\phi
-\frac{1}{2\sqrt 2}{\bar\psi}^{\nu i} \gamma_{\mu\nu\rho} \psi^{\rho j}\, \delta_{ij}
\ ,
\label{c12}
\ee
with the covariant derivative of the scalar field defined as
\be
D_\mu \phi = \partial_\mu \phi +{\sqrt 2}{\cal V}_\mu+g W_\mu\ .
\ee
Using \eq{c12} to eliminate $E_\mu$ in the Lagrangian \eq{Lag1} amounts to dualization of the 4-form potential
$E_{\mu\nu\rho\sigma}$ related to $E_\mu$ as in \eq{defFBE}\footnote{The same result is obtained by adding a total derivative Lagrange multiplier term $ eE^\mu\partial_\mu \phi$ to the Lagrangian \eq{Lag1} and integrating over $E^\mu$.}.

The shift symmetry \eq{gt} can be used to eliminate the scalar field $\phi$, by setting it to a
constant $\phi_0$. This in turn implies a compensating $\lambda =-g\eta /{\sqrt
2}$ transformation, leading to an unbroken $\U(1)$ symmetry.
Eliminating $\phi$ in this way, \eq{aux3} and \eq{c1} imply
\be
{\cal V}_\mu +\frac{1}{\sqrt 2} g W_\mu = \left(
\sigma^{-2}{\bar\psi}^i\gamma_\mu\psi^j
+\sigma{\bar\Omega}^i\gamma_\mu\Omega^j\right) \delta_{ij}\ ,
\ee
Using this equation and \eq{aux3} in the terms involving $E_\mu$ in the
action gives rise to only quartic fermion terms. The use of \eq{c1} in the
fermionic kinetic terms, however, has the effect of replacing ${\cal V}_\mu$
by $ -gW_\mu/{\sqrt 2}$, up to quartic fermion terms in the action. Thus,
altogether, the elimination of all the auxiliary fields yields, up to quartic
fermion terms, the following Lagrangian:
\bea e^{-1}\mathcal{L}_{NS}  &=& \frac{1}{2}R
-\frac{1}{2}\sigma^{-2}\partial_\mu \sigma\partial^\mu \sigma -\frac14
g^2\sigma^{-1}-\frac{1}{24}\sigma^{-2}F_{\mu\nu\rho}(B)F^{\mu\nu\rho}(B)
\nonumber\w2 &&-\frac{1}{4}\sigma F_{\mu\nu}(W)F^{\mu\nu}(W)
+\frac{1}{24} e^{-1}\varepsilon^{\mu\nu\rho\sigma\lambda\tau}
F_{\mu\nu\rho}(B)F_{\lambda\tau}(W)W_\sigma\ \nn\w2
&&-\frac{1}{2}\opsi_{\rho}\gamma^{\mu\nu\rho}{\cal D}_{\mu}\psi_{\nu}
-2\sigma^{-2}{\bar\psi}\gamma^\mu {\cal D}_\mu \psi
-2\sigma\bar{\Omega}\gamma^\mu {\cal D}_\mu \Omega \nn\w2 &&
+\sigma^{-2}\opsi_\nu\gamma^\mu\gamma^\nu \psi\partial_\mu \sigma
+\frac{g}{2\sqrt{2}}\delta^{ij}\left( \opsi_{\mu i}\gamma^{\mu}\Omega_j
+4\sigma^{-1}{\bar\Omega}_i \psi_j\right) \nn\w2 && +\frac12
F_{\mu\nu}(W) \left(\sigma
{\bar\Omega}\gamma^\rho\gamma^{\mu\nu}\psi_\rho
+2{\bar\Omega}\gamma^{\mu\nu}\psi\right)
+\frac{1}{12}F_{\mu\nu\rho}(B){\bar\Omega}\gamma^{\mu\nu\rho}\Omega
\label{RVcoupled}\w2 &&  -\frac{1}{48}\sigma^{-1}  F_{\mu\nu\rho}(B)
\Bigl( \opsi^\lambda\gamma_{[\lambda}
\gamma^{\mu\nu\rho}\gamma_{\tau]}\psi^\tau +4 \sigma^{-1}
\opsi_\lambda\gamma^{\mu\nu\rho}\gamma^\lambda\psi
-4\sigma^{-2}{\bar\psi}\gamma^{\mu\nu\rho} \psi \Bigr)\,, \nonumber
\end{eqnarray}
where
\bea
{\cal D}_\mu\psi_\nu^i &=& (\partial_\mu +
\frac14\omega_{\mu}{}^{ab}\gamma_{ab} )\psi_\nu^i +\frac{1}{2\sqrt 2}
g W_\mu\delta^{ij}\psi_{\nu j}\ , \nn\w2
{\cal D}_\mu\psi^i &=&
(\partial_\mu + \frac14\omega_{\mu}{}^{ab}\gamma_{ab})\psi^i
+\frac{1}{2\sqrt 2}g W_\mu\delta^{ij} \psi_j\ , \nn\w2 {\cal
D}_\mu\Omega^i &=& (\partial_\mu +
\frac14\omega_{\mu}{}^{ab}\gamma_{ab})\Omega^i +\frac{1}{2\sqrt 2}g
W_\mu\delta^{ij} \Omega_j\ .
\label{cdp}
\eea
This Lagrangian has the on-shell supersymmetry given, up to cubic fermion
terms, by the transformation rules for $(e_\mu^a, \psi_\mu^i,
B_{\mu\nu},\psi_i,\sigma)$ in \eq{susy1}, and for $(W_\mu,\Omega^i)$ in
\eq{susyYM1}, with the replacements
\be
Y^{ij} \ \to \ -\frac{1}{2\sqrt 2} g\sigma^{-1} \delta^{ij}\ ,\qquad {\cal
V}_\mu \ \to\ -\frac{1}{\sqrt 2} gW_\mu\ ,\qquad {\cal V}_\mu^{\prime ij}\ \to\ 0\
,\qquad \eta^i \ \to \ 0\ .
\ee
The last substitution is due to the fact that the elimination of ${\cal V}_\mu^{\prime ij}$ and $E_\mu$ in \eq{eta} gives rise to quadratic fermion terms only. These results agree with the Lagrangian obtained in \cite{Nishino:1997ff} by direct application of the Noether procedure  based on the on-shell closed supersymmetry transformations.

A dual formulation in which the field equation and Bianchi identity for the 2-form potential are interchanged is easily obtained by adding a Lagrange multiplier term
\be
\Delta{\cal L}= \frac{1}{24}\varepsilon^{\mu\nu\rho\sigma\lambda\tau} F_{\mu\nu\rho}(B)
\partial_\sigma {\widetilde B}_{\lambda\tau}\,.
\label{LM}
\ee
Treating $F_{\mu\nu\rho}(B)$ as an independent field in ${\cal L}+\Delta {\cal L}$, its field equation
can be used back in the action, yielding
\bea e^{-1}\mathcal{L}_{SS} &=&\frac{1}{2}R
-\frac{1}{2}\sigma^{-2}\partial_a\sigma\partial^a\sigma -\frac14
g^2\sigma^{-1}-\frac{1}{24}\sigma^2 G_{\mu\nu\rho}G^{\mu\nu\rho}
-\frac14\sigma F_{\mu\nu}(W)F^{\mu\nu}(W) \nn\w2 &&
-\frac12\opsi_{\rho}\gamma^{\mu\nu\rho}{\cal D}_{\mu}\psi_{\nu}
-2\sigma^{-2}{\bar\psi}\gamma^\mu {\cal D}_\mu \psi
-2\sigma\bar{\Omega}\gamma^\mu {\cal D}_\mu \Omega \nn\w2 &&
+\sigma^{-2}\opsi_\nu\gamma^\mu\gamma^\nu \psi\partial_\mu \sigma
+\frac{g}{2\sqrt{2}}\delta^{ij}\left( \opsi_{\mu i}\gamma^{\mu}\Omega_j
+4\sigma^{-1}{\bar\Omega}_i \psi_j\right) \nn\w2 && +\frac12
F_{\mu\nu}(W) \left(\sigma
{\bar\Omega}\gamma^\rho\gamma^{\mu\nu}\psi_\rho
+2{\bar\Omega}\gamma^{\mu\nu}\psi\right) -\frac12\sigma^2
G_{\mu\nu\rho}{\bar\Omega}\gamma^{\mu\nu\rho}\Omega \nn\w2 &&
+\frac18\sigma G_{\mu\nu\rho} \Bigl( \opsi^\lambda\gamma_{[\lambda}
\gamma^{\mu\nu\rho}\gamma_{\tau]}\psi^\tau -4 \sigma^{-1}
\opsi_\lambda\gamma^{\mu\nu\rho}\gamma^\lambda\psi
-4\sigma^{-2}{\bar\psi}\gamma^{\mu\nu\rho} \psi \Bigr)\ ,
\label{RVcoupled2}
\end{eqnarray}
where
\be G_{\mu\nu\rho} = 3\partial_{[\mu} {\widetilde B}_{\nu\rho]} +
3F_{[\mu\nu}(W)W_{\rho]}\ . \label{defGCS} \ee
This Lagrangian has the on-shell supersymmetry given, up to cubic fermion
terms, by the transformation rules for $(e_\mu^a, \psi_\mu^i, {\widetilde
B}_{\mu\nu},\psi_i,\sigma)$ in \eq{susy1}, and for $(W_\mu,\Omega^i)$ in
\eq{susyYM1}, with the replacements
\bea
&& Y^{ij} \ \to \ -\frac{1}{2\sqrt 2} g\sigma^{-1} \delta^{ij}\ ,\qquad {\cal
V}_\mu \ \to\ -\frac{1}{\sqrt 2} gW_\mu\ ,\qquad {\cal V}_\mu^{\prime ij}\ \to\ 0\
, \nn\w2 && B_{\mu\nu}\ \to\ {\widetilde B}_{\mu\nu}\ ,\qquad
F_{\mu\nu\rho}(B) \ \to\ \frac{1}{3!} \sigma^2
e\varepsilon_{\mu\nu\rho\sigma\lambda\tau} G^{\sigma\lambda\tau}\ ,\qquad
\eta^i \ \to \ 0\ . \label{onshell2}
\eea
These results agree with
\cite{Salam:1984cj,Nishino:1984gk,Nishino:1997ff}, after taking into account the fact that some of the fermions are
to be redefined by scaling them with a suitable power of the scalar
field $\sigma$.


\section{An Alternative Off-Shell Formulation} \label{app:totalL}


Starting from a superconformal coupling of the dilaton Weyl multiplet to
the compensating linear multiplet, we made the set of gauge choices
\eq{gf1} which led to an off-shell Poincar{\'e} supergravity with field
content \eq{m1}. If we do not insist on the canonical Einstein--Hilbert
term in the action, there exists a natural alternative set of gauge
choices given by
\be
 \boxed{\sigma=1\ ,\quad L_{ij}= \frac{1}{\sqrt 2} \delta_{ij}L\ , \quad \psi^i=0\ ,\quad b_\mu=0 }
\label{gf2}
\ee
which fix the dilatations, conformal boost and special supersymmetry, and lead to an alternative off-shell Poincar{\'e} multiplet consisting of the fields
\begin{equation}
e_{\mu}{}^a\ (15)\,,\ \ {\cal V}^\prime_{\mu}{}^{ij}\
(12)\,,\ \ {\cal V}_\mu\ (5)\,,\ \
B_{\mu\nu}\ (10)\,,\ \ L\ (1)\,,\ \ E_{\mu\nu\rho\sigma}\ (5)\,;\
\ \psi_{\mu}{}^i\ (40)\,,\ \  \varphi^i \ (8)\,.
\label{m2}
\end{equation}
Compared to the previous multiplet given in \eq{m1} $\sigma$ and
$\psi^i$  are replaced by $L$ and $\varphi^i$, and therefore this
multiplet again has $48+48$  off-shell degrees of freedom. It turns out
that this formulation of the off-shell Poincar{\'e} multiplet is very
convenient in the  construction of the only known off-shell higher
derivative invariant in $D=6$, which is a supersymmetric completion of
the Riemann tensor squared \cite{Bergshoeff:1986wc}. What makes the
gauge choice \eq{gf2} very useful in this construction is that it
furnishes a map between the off-shell supersymmetry transformations of
the Yang-Mills and Poincar{\'e} multiplets. We shall review this
construction in the next section. Here we shall focus on coupling a
vector multiplet to this alternative Poincar{\'e} supermultiplet. This
amounts to seeking an expression for ${\mathcal L}= {\mathcal L}_{R}+
{\mathcal L}_V+ g{\mathcal L}_{VL}$ in the gauge \eq{gf2}.

Starting from \eq{abelianvector} and \eq{LVLconf}, it is straightforward to obtain ${\mathcal L}_V$ and
$ g{\mathcal L}_{VL}$ in the  gauge \eq{gf2}. To construct the Einstein-Hilbert Lagrangian in this gauge, on the other hand, we first restore
superconformal invariance\footnote{To be precise, we restore
superconformal invariance partially since we do not restore the
$K$-symmetry.} by performing suitable field redefinitions in \eq{offshellPoincare}. This is achieved by replacing the fields that
transform under dilatations and special supersymmetry by
\begin{eqnarray}
{\widetilde e}_\mu{}^a &=&L^{1/4}e_\mu{}^a\ ,
\nn\\
{\widetilde\psi}_\mu^i &=& L^{1/8}\left(\psi_\mu^i
-\frac{1}{2\sqrt 2}L^{-1}\delta^{ij}\gamma_\mu\varphi_j\right)\ ,
\nn\\
{\widetilde {\cal V}}_\mu{}^{ij}&=&{\cal V}_\mu{}^{ij}-\frac{1}{\sqrt 2}L^{-1}\delta^{k(i}\bar{\varphi}_k
\psi_\mu{}^{j)}+\frac18 L^{-2}\delta^{li}\delta^{jk}\bar{\varphi}_l\gamma_{\mu}\varphi_k\,,
\nn\\
{\widetilde\sigma} &=& L^{-1/2} \sigma\ ,
\nn\w2
{\widetilde\psi}^i &=& L^{-5/8}\left( \psi^i +\frac{1}{2\sqrt 2}L^{-1}\sigma\delta^{ij}\varphi_j\right)\ ,
\nn\w2
{\widetilde E}_a &=& L^{-5/4}E_a\ ,
\nn\w2
{\widetilde\epsilon}^i&=&L^{1/8}\epsilon^i\ ,
\label{redefs1}
\end{eqnarray}
which are invariant under dilatations and special supersymmetry, as can be
checked by using the transformation rules given in
\cite{Bergshoeff:1985mz}. Next, we impose the gauge choices \eq{gf2}. Thus, we construct the Lagrangian
\begin{equation}
  {\mathcal L}_2= \Bigl( {\mathcal L}_R + {\mathcal L}_V +g{\mathcal L}_{VL} \Bigr)\Big\vert_{\sigma=1}\,,
   \label{Lagrangiansigmagauge}
\end{equation}
where ${\mathcal L}_R$ is the Lagrangian given in  (\ref{offshellPoincare})
with the field redefinitions \eq{redefs1} performed, such that the
superconformal invariance is restored, and  $\sigma=1$ refers to all the gauge choices of \eq{gf2}.
A summary of the different gauge conditions and what parts of the superconformal Lagrangian they affect can be found in table \ref{Gaugeconditions}.
\begin{center}
 \begin{table}[ht]
\caption{This table shows which gauge conditions leave which parts of the
total Lagrangian superconformal (SC) invariant and which parts not. In the
top row we have indicated on which fields the different parts of the
superconformal Lagrangian depend. \label{Gaugeconditions} } \centering
\begin{tabular}{|c|c|c|c|c|}
\hline Gauge choices & $\mathcal{L}_R$ ($L, \varphi, \sigma, \psi$) & $\mathcal{L}_{R^2}$ ($\sigma, \psi$) &
$\mathcal{L}_{\rm{V}}$ ($\sigma, \psi$) & $\mathcal{L}_{\rm{VL}}$
($L,\varphi$) \\ [0.3cm] \hline $L=1$, $\varphi^i=0$ &
breaks SC & SC & SC & breaks SC \\ [0.3cm] \hline $\sigma=1$, $\psi^i=0$ & breaks SC & breaks SC & breaks SC & SC \\ [0.3cm] \hline
\end{tabular}
\end{table}
\end{center}

Formula (\ref{Lagrangiansigmagauge}), up to quartic fermion
terms, gives rise to the following expression:
\bea
e^{-1}{\mathcal L}_2 &=& \frac12 L R +\frac12 L^{-1}\partial_\mu
L\partial^\mu L +\frac{1}{\sqrt{2}}g L \delta^{ij}Y_{ij} -\frac{1}{24} L
F_{\mu\nu\rho}(B)F^{\mu\nu\rho}(B)
\nn\\
&&  +L  {\cal V}'_\mu{}^{ij}{\cal V}^{\prime \mu}{}_{ij} -\frac14 L^{-1}E^\mu E_\mu
+\frac{1}{\sqrt 2} E^\mu \left( {\cal V}_\mu+\frac{1}{\sqrt{2}}g W_\mu
\right)
\nn\\
&& +Y^{ij} Y_{ij}-\frac{1}{4}F_{\mu\nu}(W)F^{\mu\nu}(W)
-\frac{1}{16}e^{-1}\varepsilon^{\mu\nu\rho\sigma\lambda\tau}B_{\mu\nu}F_{\rho\sigma}(W)F_{\lambda\tau}(W)
\nn\w2
&& -\frac12L \opsi_{\rho}\gamma^{\mu\nu\rho}D_{\mu}(\omega )\psi_{\nu} -{\sqrt 2}
{\bar\varphi}_i\gamma^{\mu\nu}D_\mu(\omega )\psi_{\nu j} \delta^{ij} +
L^{-1}{\bar\varphi} {\slashed D}'(\omega )\varphi -2{\bar\Omega}{\slashed D}'(\omega )\Omega
\nn\\
&& -\frac12 \left( L {\bar\psi}^\mu \gamma^\nu\psi_\nu
+{\sqrt 2} \delta_{ij}{\bar\psi}_\nu^i\gamma^\mu\gamma^\nu\varphi^j\right)
L^{-1}\partial_\mu L -\frac{1}{\sqrt 2}g L
{\bar\Omega}_i\gamma^\mu\psi_{\mu j} \delta^{ij} \nn\w2 && +2 g
{\bar\Omega}\varphi +\frac12{\bar\Omega}\gamma^\mu\gamma\cdot
F(W)\psi_\mu +\frac{1}{12}{\bar\Omega}\gamma\cdot F(B)\Omega
+\frac{1}{24} L^{-1}{\bar\varphi}\gamma\cdot F(B)\varphi
\nn\\
&&  -\frac{1}{48}L F_{\mu\nu\rho}(B) \left(\opsi^\lambda\gamma_{[\lambda}
\gamma^{\mu\nu\rho}\gamma_{\tau]}\psi^\tau +2{\sqrt 2}
L^{-1}{\bar\psi}_{\lambda i}\gamma^{\lambda\mu\nu\rho}\varphi_j \delta^{ij}\right)
\nn\\
&& -\frac{1}{4\sqrt 2} E_\rho \left(\bar \psi_\mu^i
\gamma^{\rho\mu\nu}\psi_\nu^j \delta_{ij}-2{\sqrt 2} L^{-1}
{\bar\psi}_\sigma \gamma^\rho\gamma ^\sigma\varphi +2
L^{-2}{\bar\varphi}_i\gamma^\rho\varphi_j\delta ^{ij}\right)
\nn\w2
&& +\frac12 {\cal V}^{\prime \mu ij}\left( {2\sqrt
2}{\bar\varphi}^k\psi_{\mu i}\delta_{jk}
-3L^{-1}{\bar\varphi}_i\gamma_\mu\varphi_j\right)\ ,
\label{Lag2} \eea
where $E_\mu$ is not an independent field but rather the dual of the field strength for the four-form potential, see \eq{defFBE}, the derivative
$D_\mu(\omega )\psi_\nu$ is $\U(1)$ covariant as in \eq{cd1}, and the derivatives  $D'_\mu(\omega )\varphi$ and
$D'_\mu(\omega )\Omega$ are $\SU(2)$ covariant as in \eq{cd2}.

The off-shell supersymmetry transformations for this Lagrangian are to
be obtained from those of the  dilaton Weyl multiplet upon fixing the
gauges \eq{gf2}. It is important to note that the field redefinitions
\eq{redefs1} are not to be performed in this process since these
transformations are independent of the linear multiplet fields that were
used to impose the gauge choices \eq{gf1}. In obtaining these
transformations, the compensating transformations required to maintain
the gauge (\ref{gf2}) must also be incorporated. These are a
compensating special supersymmetry transformation and a compensating
(traceless) $\SU(2)$ transformation with parameters given by (up to
cubic fermion terms)
\bea
\eta^i&=&\frac{1}{48} \gamma\cdot F(B) \epsilon^i\ ,
\nn\\
\lambda'{}^{ij}&=&-\frac{1}{\sqrt{2}L}\Bigl(S'{}^{k(i}\delta^{j)l}\epsilon_{kl}\Bigr)\ ,
\label{decompositionlaw}
\eea
where\footnote{It is instructive to write out the $\lambda'$ parameter in components:
\[
\lambda'{}^{11}=-\lambda'{}^{22}=\frac{1}{\sqrt{2}L}S'{}^{21}\,, \qquad  \lambda'{}^{12}=-\frac{1}{\sqrt{2}L}S'{}^{11}\,.
\]
}
\be
S'{}^{ij}\equiv\bar{\varepsilon}^{(i}\varphi^{j)}-\frac{1}{2}\delta^{ij}\bar{\varepsilon}^{k}\varphi^\ell \delta_{k\ell}
\ee
is the supersymmetry transformation of the traceless part of $L^{ij}$.
Note that the prime stands for `traceless', i.e. $S'{}^{ij}\delta_{ij}=0$.
These compensating transformations can be obtained from the transformation rules for $\psi^i$ and $L^{ij}$ given in \cite{Bergshoeff:1985mz}.

Thus, using the supersymmetry transformation rules
for the dilaton Weyl multiplet provided in \cite{Bergshoeff:1985mz,Coomans:2011ih}, the
gauge conditions \eq{gf2} and the compensating transformations with parameters given in (\ref{decompositionlaw}), we find that the supersymmetry
transformations of the off-shell Poincar{\'e} multiplet, up to cubic fermion terms, take the form
\bea
\delta e_{\mu}{}^a&=&\frac12{\bar\epsilon}\gamma^a\psi_{\mu}\ ,
\nn\\
\delta \psi_{\mu}{}^i&=& (\partial_{\mu} +\frac14\omega_{\mu
ab}\gamma^{ab})\epsilon^i +{\cal V}_\mu{}^i{}_j\epsilon^j +\frac18
F_{\mu\nu\rho}(B)\gamma^{\nu\rho}\epsilon^i\ ,
\nn\\
\delta B_{\mu\nu}&=&-{\bar\epsilon}\gamma_{[\mu}\psi_{\nu]}\ ,
\nn\\
\delta \varphi^i &=& \frac{1}{2\sqrt 2} \gamma^\mu
\delta^{ij}\partial_\mu L \epsilon_j -\frac14 \gamma^\mu E_\mu\epsilon^i
+\frac{1}{\sqrt2}\gamma^\mu {\cal V}'{}_\mu^{(i}{}_k \delta^{j)k} L
\epsilon_j - \frac{1}{12\sqrt2}L\delta^{ij}\gamma\cdot F(B) \epsilon_j \
, \nn\w2 \delta L &=& \frac{1}{\sqrt 2} {\bar\epsilon}^i
\varphi^j\delta_{ij} \ , \nn\w2 \delta E_{\mu\nu\rho\sigma} &=&
L{\bar\epsilon}^i\gamma_{[\mu\nu\rho}\psi_{\sigma]}^j\delta_{ij}
-\frac{1}{2\sqrt 2} {\bar\epsilon}\gamma_{\mu\nu\rho\sigma}\varphi \ ,
\nn\w2
\delta {\cal
V}_{\mu}&=&\frac{1}{2}\bar{\epsilon}^{i}\gamma^{\nu }\widehat{R}_{\mu\nu }{}^{j}(Q)\delta_{ij}
+\frac{1}{12}\bar{\epsilon}^{i}\gamma\cdot
F(B)\psi_{\mu}{}^{j}\delta_{ij}-2\lambda'{}^{i}{}_k{\cal V}_{\mu}'{}^{jk}\delta_{ij} \ ,
\nn\w2
\delta {\cal
V}_{\mu}'{}^{ij}&=&\frac{1}{2}\bar{\epsilon}^{(i}\gamma^{\nu }\widehat{R}_{\mu\nu }{}^{j)}(Q)
+\frac{1}{12}\bar{\epsilon}^{(i}\gamma\cdot
F(B)\psi_{\mu}{}^{j)}-\frac{1}{4}\bar{\epsilon}^{k}\gamma^{\nu }\widehat{R}_{\mu\nu }{}^\ell (Q)\delta_{k\ell }\delta^{ij} \nonumber \\
&&-\frac{1}{24}\bar{\epsilon}^{k}\gamma\cdot F(B)\psi_{\mu}{}^\ell \delta_{k\ell }\delta^{ij}+\partial_{\mu}\lambda'{}^{ij}-\lambda'{}^{(i}{}_k\delta^{j)k}{\cal V}_{\mu}\,,
\label{susy2}
\eea
where
\begin{equation}
  \widehat{R}_{\mu\nu }{}^i(Q)=  2D_{[\mu }(\omega )\psi _{\nu  ]}^i
  -2{\cal V}_{[\mu }^{\prime\,ij}\psi _{\nu  ]j}
  +\frac{1}{4}\gamma ^{ab}\psi _{[\nu }F_{\mu ]ab}\,.
 \label{RQ}
\end{equation}
The supersymmetry transformations of the off-shell vector multiplet are (up to cubic fermion terms)
\begin{eqnarray}
\delta W_{\mu}&=&-\bar{\epsilon}\gamma_{\mu}\Omega\ ,
\nonumber\\
\delta \Omega^{i}&=&\frac{1}{8}\gamma^{\mu\nu}F_{\mu\nu}\epsilon^i-\frac{1}{2}Y^{ij}\epsilon_j\ ,
\nonumber\\
\delta Y^{ij} &=& -\bar{\epsilon}^{(i}\gamma^{\mu}D'_{\mu}(\omega )\Omega^{j)}
+\frac18 {\bar\epsilon}^{(i} \gamma^\mu\gamma\cdot F(B) \psi_\mu^{j)}
-\frac{1}{24} {\bar\epsilon }^{(i} \gamma\cdot F(B) \Omega^{j)} \nonumber\\
&&-\frac12 Y^{k(i} {\bar\epsilon}^{j)} \gamma^\mu \psi_{\mu k}-2\lambda'{}^{(i}{}_kY^{j)k}\ .
\label{susy3}
\end{eqnarray}
To keep the notation relatively simple we did not use the explicit expression for $\lambda'{}^{ij}$ in the above transformation rules. Remember that it is given in (\ref{decompositionlaw}).

Considering the Lagrangian \eq{Lag2} by itself, that is, without any higher derivative extension, all the auxiliary fields,
namely $({\cal V}_\mu^{ij}, E_{\mu\nu\rho\sigma}, Y^{ij})$ can be eliminated, thereby arriving at the on-shell formulation.
Computations similar to those described in detail in section \ref{ss:elimAuxF} imply that the on-shell Lagrangian, up to quartic fermion terms, is obtained from \eq{Lag2} by the following substitutions:
\be
Y^{ij} \ \to \ -\frac{1}{2\sqrt 2} g \delta^{ij}L\ ,\quad {\cal
V}_\mu \ \to\ -\frac{1}{\sqrt 2} gW_\mu\ ,
\quad {\cal V}_\mu^{\,\prime ij}\ \to\ 0 \ ,\quad E_\mu\ \to\ 0\ .
\label{onshell3}
\ee
The on-shell supersymmetry transformations, up to cubic fermion terms,  are obtained from \eq{susy2} and \eq{susy3} by making these substitutions, and dropping the transformation rules for the auxiliary fields
$(E_{\mu\nu\rho\sigma}, {\cal V}_\mu^{ij}, Y^{ij})$.


\section{Inclusion of the \texorpdfstring{$R_{\mu\nu ab}R^{\mu\nu ab}$}{R-squared} invariant}
\label{Section:gaugingwithR2terms}

In this section we add an off-shell supersymmetric Riemann tensor squared
term to the Lagrangian ${\mathcal L}_2$, defined in (\ref{Lagrangiansigmagauge}),
which we constructed in the  gauge \eq{gf2}. This gauge gave rise to an alternative off-shell
formulation of the Poincar{\'e} multiplet. In the first subsection we begin with a
review of the construction of the Riemann squared invariant
\cite{Bergshoeff:1986wc}. In the second subsection we consider the total Lagrangian and briefly discuss the gauging procedure and the elimination of auxiliary fields.

\subsection{Construction of the \texorpdfstring{$R_{\mu\nu ab}R^{\mu\nu ab}$}{R-squared} invariant}
To begin with, we
shall review a map between the Yang-Mills supermultiplet and a set of fields
in the alternative Poincar{\'e} multiplet discussed in the previous section.  We follow the  discussion in
\cite{Bergshoeff:1987rb}. This map can be used, together with an expression
for the superconformal action for the Yang-Mills multiplet given in
\cite{Bergshoeff:1985mz},  to write down a supersymmetric Riemann tensor
squared action. We will describe this in detail below.

In establishing the map between the Yang-Mills and Poincar{\'e} multiplets, it is
important to consider the full supersymmetry transformations, including the
cubic fermion terms which have been omitted so far. In particular, this means
that we need to keep track of the complete spin connection, containing the
fermionic torsion terms. This is due to the fact that, while the fermionic
torsion gave rise to only quartic fermion terms in the Lagrangians considered
above, in the case of the Riemann tensor square invariant under consideration in
this section, the same fermionic torsion will contribute to terms that are bilinear in the fermion
terms. We shall show this explicitly below. In the following, we shall need the
(full) supersymmetry transformation rules only for the fields
$(e_\mu^a,\psi_\mu, {\cal V}_\mu^{ij}, B_{\mu\nu})$, and the Yang-Mills
multiplet fields $(W_\mu^I, \Omega^I, Y^{ij I})$, where $I$ labels the
adjoint representation of the Yang-Mills gauge group.

We begin with the supersymmetry transformation rules of $(e_\mu^a,
\psi_\mu, {\cal V}_\mu^{ij},B_{\mu\nu})$ in the gauge \eq{gf2}. Up to
cubic fermions the transformation rules are already given in \eq{susy2}.
In this section we will, however, keep the complete $\SU(2)$ symmetry,
i.e.~we do not impose $L^{ij}=\frac{1}{\sqrt{2}}L\delta^{ij}$. In this way we
do not need to accommodate for the compensating $\SU(2)$ transformations
proportional to $\lambda'$ in (\ref{susy2}).\footnote{In this section we
only want to establish a map between the Poincar{\'e} multiplet and the
Yang-Mills multiplet and propose an $R^2$-invariant based on the action
for the Yang-Mills multiplet. Both actions are invariant under the
$\SU(2)$ R-symmetry. To prove the validity of this map, we need the full
nonlinear SUSY transformation rules. After we construct the action we
can still impose the gauge $L^{ij}=\frac{1}{\sqrt{2}}L\delta^{ij}$.
This will not affect the $R^2$-invariant. It modifies the supersymmetry transformation rules with
$\SU(2)$ compensating transformations, which leave the action separately invariant. The resulting transformations are those given already
in (\ref{susy2}). }
The full version of the supersymmetry transformations is given by \cite{Bergshoeff:1986wc}
%
%
\begin{eqnarray}
\delta e_\mu{}^a &=&\frac12{\bar\epsilon}\gamma^a\psi_\mu\,, \nonumber \\
\delta \psi_{\mu}{}^i&=&\partial_{\mu}\epsilon^i+\frac{1}{4}{\widehat\omega}_{+\mu}{}^{ab}\gamma_{ab}\epsilon^i
+{\cal V}_{\mu}{}^i{}_j\epsilon^j\,\equiv\, D_\mu({\widehat\omega}_+)\epsilon^i+{\cal V}_{\mu\ j} ^{\,\prime\, i}\epsilon ^j\ , \nonumber \\
\delta {\cal V}_{\mu}{}^{ij}&=&-\frac{1}{2}\bar{\epsilon}^{(i}\gamma^\lambda {\widehat R}_{\lambda\mu}{}^{j)}(Q)+\frac{1}{12}\bar{\epsilon}^{(i}\gamma\cdot {\widehat F}(B) \psi_{\mu}{}^{j)}\,, \nonumber \\
\delta B_{\mu\nu}&=&-\bar{\epsilon}\gamma_{[\mu}\psi_{\nu]}\,, \label{fullsusy}
\end{eqnarray}
where the fermionic torsion and the different supercovariant objects are defined as
%
\begin{eqnarray}
{\widehat\omega}_{\mu \pm}{}^{ab} &=& {\widehat\omega}_{\mu}{}^{ab} \pm \frac{1}{2}\widehat{F}_{\mu}{}^{ab}(B)\,, \nonumber \\
{\widehat\omega}_{\mu}{}^{ab}&=&2e^{\nu [a}\partial_{[\mu}e_{\nu]}{}^{b]}-e^{\rho[a}e^{b]\sigma}e_{\mu}{}^{c}\partial_{\rho}e_{\sigma c}
+ K_\mu{}^{ab}\,,\nonumber\\
K_\mu{}^{ab}&=& \frac{1}{4}\bigl(2\bar{\psi}_{\mu}\gamma^{[a}\psi^{b]}+\bar{\psi}^a\gamma_{\mu}\psi^b\bigr)\,, \nonumber \\
\widehat{F}_{\mu\nu\rho}(B)&=&3\partial_{[\mu}B_{\nu\rho]}+\frac{3}{2}\opsi_{[\mu}\gamma_{\nu}\psi_{\rho]}\,, \nonumber \\
{\widehat R}_{\mu\nu}{}^i(Q) &=& 2\left(\partial_{[\mu}+\frac{1}{4}{\widehat\omega}_+{}_{[\mu}{}^{ab}\gamma_{ab}\right)\psi_{\nu]}{}^i+2{\cal V}_{[\mu}{}^i{}_j\psi_{\nu]}{}^j\,. \label{torsionful}
\end{eqnarray}
Next, we consider the following transformations \cite{Bergshoeff:1987rb}
\begin{eqnarray}
\delta {\widehat\omega}_{-\mu}{}^{ab}&=&-\frac{1}{2}\bar{\epsilon}\gamma_\mu {\widehat R}^{ab}(Q)\,, \nonumber \\
\delta {\widehat R}^{abi}(Q) &=&\frac{1}{4}\gamma^{cd}\epsilon^i\widehat{R}_{cd}{}^{ab}({\widehat\omega}_-)-\widehat{F}^{abij}({\cal V})\epsilon_j\,, \nonumber \\
\delta \widehat{F}^{abij}({\cal V})&=&-\frac{1}{2}\bar{\epsilon}^{(i}\gamma^{\mu}{\widehat D}_{\mu}{\widehat R}^{abj)}(Q)+\frac{1}{48}\bar{\epsilon}^{(i}\gamma\cdot {\widehat F}(B) {\widehat R}^{abj)}(Q)\,,
\label{omegavector}
\end{eqnarray}
where ${\widehat F}_{\mu\nu}{}^{ij}({\cal V})$ and
${\widehat R}_{\mu\nu}{}^{ab}({\widehat\omega}_-)$ are the supercovariant curvatures of
${\cal V}_{\mu}{}^{ij}$ and ${\widehat\omega}_{-\,\mu}{}^{ab}$, respectively:
\begin{eqnarray}
{\widehat F}_{\mu\nu}{}^{ij}({\cal V})&=&F_{\mu\nu}{}^{ij}({\cal V})-\opsi_{[\mu}{}^{(i}\gamma^\rho {\widehat R}_{\nu]\rho }{}^{j)}(Q)-\frac{1}{12}\opsi_{[\mu}{}^{(i}\gamma \cdot {\widehat F}(B)\psi_{\nu]}{}^{j)}\,, \nonumber \\
\widehat{R}_{\mu\nu}{}^{ab}({\widehat\omega}^-)&=&R_{\mu\nu}{}^{ab}({\widehat\omega}^-)
+\opsi_{[\mu}\gamma_{\nu]}{\widehat R}^{ab}(Q)\,,
\nonumber\\
{\widehat D}_{\mu} {\widehat R}^{abi}(Q)&=&\partial_{\mu}{\widehat R}^{abi}(Q)
+\frac{1}{4}{\widehat\omega}_{\mu}{}^{cd}\gamma_{cd} {\widehat R}^{abi}(Q)+{\cal V}_{\mu}{}^i{}_j {\widehat R}^{abj}(Q) \nonumber \\
&&-\frac{1}{4}\gamma^{cd}\psi_{\mu}{}^i{\widehat R}_{cd}{}^{ab}({\widehat\omega}_-)+{\widehat F}^{abij}({\cal V})\psi_{\mu j}+2{\widehat\omega}_{-\mu}{}^{[ac} {\widehat R}_c{}^{b]i}(Q)\,.
\end{eqnarray}
We now compare the above transformation rules with those of the ${\cal N}=(1,0)$,
$D=6$ vector multiplet  \cite{Bergshoeff:1985mz}
\begin{eqnarray}
\delta W_{\mu}{}^I&=&-\bar{\epsilon}\gamma_{\mu}\Omega^I\,, \nonumber \\
\delta \Omega^{Ii}&=&\frac{1}{8}\gamma\cdot{\widehat F}^I(W)\epsilon^i-\frac{1}{2}Y^{Iij}\epsilon_j\,, \nonumber \\
\delta Y^{Iij} &=& -\bar{\epsilon}^{(i}\gamma^{\mu}\widehat{D}_{\mu}\Omega^{j)I}
+\frac{1}{24}\bar{\epsilon}^{(i}\gamma\cdot {\widehat F}(B) \Omega^{j)I}\,, \label{vector}
\end{eqnarray}
where
\begin{eqnarray}
{\widehat F}_{\mu\nu}{}^I(W)&=& F_{\mu\nu}{}^I(W) +2\opsi_{[\mu}\gamma_{\nu]}\Omega^I\,,\nn\w2
{\widehat D}_{\mu}\Omega^{Ii}&=&\partial_{\mu}\Omega^{Ii}+\frac{1}{4}{\widehat\omega}_{\mu}{}^{ab}\gamma_{ab}\Omega^{Ii}+V_{\mu}{}^i{}_j\Omega^{Ij} \nonumber \\
&&-\frac{1}{8}\gamma \cdot {\widehat F}^I(W) \psi_{\mu}{}^i+\frac{1}{2}Y^{I\,ij}\psi_{\mu j}-f_{KL}{}^IW_{\mu}{}^K\Omega^{Li}\,.
\end{eqnarray}
We observe that the transformation rules (\ref{omegavector}) and
(\ref{vector}) become identical by making the following identifications:
\begin{equation}
\Bigl(-2{\widehat\omega}_{-\mu}{}^{ab},-{\widehat R}^{abi}(Q),-2{\widehat F}^{abij}({\cal V})\Bigr) \longrightarrow \Bigl(W_{\mu}{}^I,\Omega^{Ii},Y^{Iij}\Bigr)\,.
\label{map}
\end{equation}
Using this observation we can now easily write down a supersymmetric
$R^2$-action using the superconformal invariant exact action formula for the Yang-Mills multiplet constructed in
\cite{Bergshoeff:1985mz}. In the gauge \eq{gf2} and up to quartic fermions, the Lagrangian becomes
\begin{eqnarray}
\left.e^{-1}\mathcal{L}_{\rm YM}\right|_{\sigma =1}&=&-\frac{1}{4}F_{\mu\nu}{}^I(W)F^{\mu\nu I}(W)
-2\bar{\Omega}^I \gamma^\mu D'_\mu(\omega) \Omega^I +Y^{Iij}Y_{ij}^I
+\frac{1}{12}F_{\mu\nu\rho}(B) \bar{\Omega}^I\gamma^{\mu\nu\rho} \Omega^I \nonumber \\
&&-\frac{1}{16}e^{-1}\varepsilon^{\mu\nu\rho\sigma\lambda\tau}B_{\mu\nu}F_{\rho\sigma}^I(W) F_{\lambda\tau}^I(W)+\frac12 F_{\nu\rho}{}^I {\bar\Omega}^I\gamma^\mu\gamma^{\nu\rho}\psi_\mu\,.
\label{actionformula}
\end{eqnarray}
Using the map \eq{map}
in this formula produces the result for the supersymmetrized Riemann
tensor squared action. In presenting the results up to quartic fermion
terms, it is useful to note the following simplification in the
torsionful spin connection
\bea
{\widehat\omega}_{\mu-}{}^{ab} &=& \omega_{\mu +}{}^{ab} +\frac12 {\bar\psi}^a\gamma_\mu\psi^b\ ,
\nn\\
\omega_{\mu\pm}{}^{ab} &\equiv& \omega_\mu{}^{ab}\pm \frac12
F_\mu{}^{ab}(B)\ ,
\label{spinconnbostors}
\eea
where $\omega_\mu{}^{ab}$ is the standard torsion-free connection. The
map \eq{map} applied to the action formula \eq{actionformula} then
yields, up to quartic fermion terms, the result
\footnote{To obtain \eq{R2} we used $-\mathcal{L}_{\rm{V}}$. Note also that
$F_{\mu\nu}{}^{ij}({\cal V})=\frac12 F_{\mu\nu}({\cal V})\delta^{ij}+F'_{\mu\nu}{}^{ij}({\cal V})$
where $F_{\mu\nu}({\cal V})=2\partial_{[\mu}{\cal V}_{\nu]} +2{\cal V}'_{\mu}{}^{i}{}_k {\cal V}'_\nu{}^{jk}\,\delta_{ij}$ and $F'_{\mu\nu}{}^{ij}({\cal V})
=2\partial_{[\mu} {\cal V}'_{\nu]}{}^{ij}
-2\delta^{k(i} \,{\cal V}_{[\mu} {\cal V}^{\prime\, j)}_{\nu]\ k}$.}
\begin{eqnarray}
\left.e^{-1} \mathcal{L}_{\rm{R^2}}\right|_{\sigma =1}&=&R_{\mu\nu}{}^{ab}(\omega_-)R^{\mu\nu}{}_{ab}(\omega_-)-2F^{ab}({\cal V})F_{ab}({\cal V})-4F^{\prime abij}({\cal V})F'_{abij}({\cal V})
\nonumber\\
&& +\frac{1}{4}e^{-1}\varepsilon^{\mu\nu\rho\sigma\lambda\tau}B_{\mu\nu}R_{\rho\sigma}{}^{ ab}(\omega_-)R_{\lambda\tau}{}_{ab}(\omega_-)
 \nonumber \\
&& +2\bar{R}_{+ab}(Q)\gamma^\mu D_\mu (\omega,\omega_-)R_{+}^{ab}(Q)-R_{\nu\rho}{}^{ab}(\omega_-)\bar{R}_{+ab}(Q)\gamma^{\mu}\gamma^{\nu\rho}\psi_{\mu} \nonumber \\
&& -8F'_{\mu\nu}{}^{ij}({\cal V})\left({\bar\psi}^\mu_i \gamma_\lambda R^{\lambda\nu}_{+j}(Q)
+\frac16 {\bar\psi}^\mu_i \gamma\cdot F(B)\psi^\nu_j\right) \nonumber \\
&&-\frac{1}{12}{\bar R}_{+}^{ab}(Q)\gamma\cdot F(B) R_{+ab}(Q)
 \nonumber\\
&&-\frac12 \Bigl[ D_\mu(\omega_-,\Gamma_+) R^{\mu\rho ab} (\omega_-)
-2 F_{\mu\nu}{}^\rho(B) R^{\mu\nu ab} (\omega_-)\Bigr] {\bar\psi}_a\gamma_\rho\psi_b\,,
\label{R2}
\end{eqnarray}
where
\bea
D_\mu (\omega,\omega_-)R_{+}^{ab i}(Q) &=&
(\partial_\mu +\frac14\omega_\mu{}^{cd}\gamma_{cd} )R_{+}^{ab i}(Q)
-2 \omega_{\mu -}{}^{c[a} R_{+c}{}^{b]i}(Q)  +{\cal V}_\mu{}^i{}_j R^{ab\,j}_{+}(Q)\ ,
\nn\w2
R_{+\mu\nu}{}^i(Q) &=& 2D_{[\mu}(\omega_+)\psi_{\nu]}^i-2{\cal V}'_{[\mu }{}^{ij}\psi _{\nu ]j}\ ,
\label{cd3}
\eea
and the torsionful modification of the Christoffel symbol $\Gamma^\rho_{\mu\nu\pm}$
is defined as
\be
\Gamma^\rho_{\mu\nu\pm} \equiv \Gamma^\rho_{\mu\nu} \pm \frac12F_{\mu\nu}{}^\rho(B)\,.
\label{chrsymbbostors}
\ee
This completes the construction of the supersymmetric $R^2$-invariant.

\subsection{The Total Gauged \texorpdfstring{$R+R^2$}{R+R2} Supergravity Lagrangian}

We now want to discuss what the influence is of these $R^2$-terms on the
gauging procedure described in section \ref{sect:Rsymmgauging}. The
Lagrangian we consider is the following
\begin{equation}
\mathcal{L}_{\rm{total}}=\mathcal{L}_2
-\left.\frac{1}{8M^2}\mathcal{L}_{R^2}\right|_{\sigma =1}\,, \label{totalL}
\end{equation}
with ${\mathcal L}_2$ given in \eq{Lag2} and $\mathcal{L}_{R^2}$ given
in \eq{R2} and with $M$ an arbitrary mass parameter. Recall that
${\mathcal L}_2$ has been obtained as a sum of off-shell supersymmetric
Lagrangians ${\mathcal L}_R, {\mathcal L}_V$ and ${\mathcal L}_{VL}$ and
that ${\mathcal L}_{R^2}$ is off-shell supersymmetric as well. Thus all
four parts of the total Lagrangian we consider are \textit{completely
off-shell supersymmetric.} So their sum, the total Lagrangian, is still
off-shell supersymmetric. In particular, the bosonic part of this total
Lagrangian, which will be the starting point of the next section, takes
the form
\begin{eqnarray}
e^{-1}\mathcal{L}_{\rm bos}^{\rm tot}&=& \frac{1}{2}LR+\frac{1}{\sqrt{2}} gL\delta^{ij}Y_{ij}
+Y^{ij}Y_{ij}+\frac{1}{2}L^{-1}\partial_{\mu}L\partial^{\mu}L-\frac{1}{24} L F_{\mu\nu\rho}(B)F^{\mu\nu\rho}(B) \nonumber\\
&&+2LZ_\mu Z^{*\mu}-\frac{1}{4}L^{-1}E_\mu  E^\mu
+\frac{1}{\sqrt{2}}E^{\mu}\bigl({\cal V}_{\mu}+\frac{1}{\sqrt{2}}gW_{\mu}\bigr) \nonumber\\
                         &&-\frac{1}{4}F_{\mu\nu}(W)F^{\mu\nu}(W)-\frac{1}{16}e^{-1}\varepsilon^{\mu\nu\rho\sigma\lambda\tau}B_{\mu\nu}
                         F_{\rho\sigma}(W)F_{\lambda\tau}(W) \nonumber\\
&&-\frac{1}{8M^2}\Bigl[ R_{\mu\nu}{}^{ab}(\omega_-)R^{\mu\nu}{}_{ab}(\omega_-)
-2F^{\mu\nu}({\cal V})F_{\mu\nu}({\cal V})-8F^{\mu\nu}(Z)F^*_{\mu\nu}(Z) \nonumber\\
&&+\frac{1}{4}e^{-1}\varepsilon^{\mu\nu\rho\sigma\lambda\tau}B_{\mu\nu}
R_{\rho\sigma}{}^ {ab}(\omega_-)R_{\lambda\tau}{}_{ab}(\omega_-)\Bigr]\ ,
 \label{totalbos}
\end{eqnarray}
where we have defined the complex vector fields
\be
Z_\mu \equiv {\cal V'}_\mu^{11} +\rmi{\cal V'}_\mu^{12}\ , \qquad
Z_\mu^* ={\cal V'}_{\mu 11} -\rmi{\cal V'}_{\mu 12}=-{\cal V'}_\mu ^{11} +\rmi{\cal V'}_\mu ^{12}\ ,
\ee
and field strengths
\be
F_{\mu\nu}({\cal V}) =2\partial_{[\mu} {\cal V}_{\nu]} -4\rmi Z_{[\mu} {Z}^*_{\nu]}\ ,
\qquad
F_{\mu\nu}(Z) = 2\partial_{[\mu} Z_{\nu]} -2\rmi{\cal V}_{[\mu} Z_{\nu]}\,.
\label{cf}
\ee
The part of the total Lagrangian containing the fermions is given in \eq{Lag2} and (\ref{R2}). 
None of the auxiliary fields have been eliminated so far, and the
Lagrangian still possesses the $\U(1)_R\times \U(1)$ symmetry. The field
equations for the auxiliary fields $Z_\mu$ and ${\cal
V}_\mu$ are not algebraic anymore and therefore they become propagating.
The auxiliary fields $(Y_{ij}, E_{\mu\nu\rho\sigma})$, on the other
hand, still have algebraic field equations. Their elimination, as well
as the breaking of $\U(1)_R \times \U(1)$ down to a single $\U(1)$ will
be discussed in the next section.

At this point one may pursue two different lines of thought. The first
is to consider the theory as a toy model in its own right and consider
$M^2$ as an arbitrary (not necessarily large) parameter of the
theory. The other is to think of $|M^2|$ as being large compared to a cut-off $\Lambda$ in the momentum squared.
In that case
the theory is to be treated as an effective  field theory that describes phenomena with external momenta not exceeding
${\sqrt \Lambda}$. Furthermore, the curvature-squared term is a correction term of order
$ \Lambda/|M^2|$.\footnote{In this case, the ghosts expected to arise in the spectrum will have masses of order $|M| >> \Lambda$ which can be ignored in the effective theory valid up to the energy scale
$\Lambda$.}
In this case we can compare the theory with an effective
(up to curvature squared terms) string theory Lagrangian
compactified to 6 dimensions. In the next section we will only focus on
the first line of thought. Let us however briefly comment on the
elimination of the $Z_\mu$ and ${\cal V}_\mu$. For $\Lambda/|M^2|
\ll 1$, one particular consequence of eliminating the auxiliary fields
up to order $\Lambda/|M^2|$ is that
\begin{equation}
{\cal V}^{\mu}=-\frac{1}{\sqrt{2}}\left( g W^{\mu}+\frac{L^{-1}}{M^2}\nabla_{\nu}\Bigl(F^{\mu\nu}({\cal V})+\cdots \right) =0\,, \label{EOM1}
\end{equation}
which, upon substitution back into the Lagrangian \eq{totalbos}, and trivial elimination of the
other auxiliary fields, gives
\bea
e^{-1}{\mathcal L}_{\rm bos}^{\rm tot}&=& \frac{1}{2}LR-\frac{1}{4}g^2L^2
+\frac{1}{2}L^{-1}\partial_{\mu}L\partial^{\mu}L-\frac{1}{24} L F_{\mu\nu\rho}(B)F^{\mu\nu\rho}(B)
\nonumber \\
&&-\frac14 \left(1-\frac{g^2}{2M^2}\right)F_{\mu\nu}(W)F^{\mu\nu}(W)
-\frac{1}{16}e^{-1}\varepsilon^{\mu\nu\rho\sigma\lambda\tau}B_{\mu\nu}
F_{\rho\sigma}(W) F_{\lambda\tau}(W)
\nonumber \\
&&-\frac{1}{8M^2}\Bigl[ R_{\mu\nu}{}^{ab}(\omega_-)R^{\mu\nu}{}_{ab}(\omega_-)
+\frac14 e^{-1}\varepsilon^{\mu\nu\rho\sigma\lambda\tau}B_{\mu\nu}
R_{\rho\sigma}{}^{ab}(\omega_-)R_{\lambda\tau}{}_{ab}(\omega_-)\Bigr]\,.
\label{totalbos2}
\eea
We observe that $g^2=2M^2$ is a critical
coupling at which the Maxwell kinetic term  drops out. However, this is a regime for large coupling constant, and as such it falls outside the regime of
perturbative validity. We shall nonetheless examine further what happens for this coupling in the next section where we study the field
equations in more detail. Another property of this Lagrangian is that
the dualization of the 2-form potential by adding the Lagrange
multiplier term \eq{LM} and integrating over $F(B)$, gives a dualized
field strength of the form \eq{defGCS} which now contains also a Lorentz
Chern--Simons term.

In the Lagrangian \eq{totalbos} presented above, the Einstein--Hilbert
term is not in a canonical frame. The metric can be rescaled
appropriately to obtain the canonical Einstein--Hilbert action, still
remaining in the formulation in terms of the off-shell Poincar{\'e}
supermultiplet displayed in \eq{m2}. Alternatively, we can employ the
off-shell Poincar{\'e} multiplet that results from the gauge choices
(\ref{gf1}) by following the following procedure. Since the Lagrangian
${\mathcal L}_1$ given in \eq{Lag1} is already formulated in the desired
supermultiplet formulation, we need to only construct ${\mathcal L}_{\rm
R^2}$ in the same gauge. This can be done as follows. Firstly, we
restore the superconformal invariance (again modulo the conformal boosts
which do not affect the final result) in \eq{R2} by going over to hatted fields
defined by
\begin{eqnarray}
{\widehat e}_{\mu}{}^a &=&\sigma^{1/2}e_{\mu}{}^a\ , \nn\w2
{\widehat\psi}_{\mu}{}^i &=&
\sigma^{1/4}\psi_{\mu}{}^i+\sigma^{-3/4}\gamma_{\mu}\psi^i\ , \nn\w2 {\widehat
{\cal V}}_{\mu}{}^{ij} &=& {\cal
V}_{\mu}{}^{ij}-4\sigma^{-1}\opsi^{(i}\psi_{\mu}{}^{j)}
-4\sigma^{-2}\opsi^{(i}\gamma_{\mu}\psi^{j)}\
, \nn\w2 {\widehat L} &=& \sigma^{-2}L\ , \nn\w2 {\widehat\varphi}^i &=&
\sigma^{-9/4}\left( \varphi^i -2{\sqrt
2}\sigma^{-1}L\delta^{ij}\psi_j\right)\ , \nn\w2
{\widehat Y}_{ij} &=&
\sigma^{-1}\left( Y_{ij}+\frac13
{\bar\psi}^\mu_{(i}\gamma_\mu\Omega_{j)}\right)\ , \nn\w2 {\widehat\Omega}^i
&=& \sigma^{-3/4}\Omega^i\ , \nn\w2
{\widehat\epsilon}^i &=&
\sigma^{1/4}\epsilon^i\ . \label{hattedfields}
\end{eqnarray}
%
Next, we impose the gauge conditions listed in \eq{gf1} and add the result to \eq{Lag1}
to obtain the full $R+R^2$ theory in this gauge. This straightforward
computation will not be carried out here since we shall be working in the gauge \eq{gf2} which leads to the result \eq{totalL} for the total Lagrangian.


\section{Vacuum Solutions}\label{ss:vacuumsol}


The purpose of this section is to investigate the different
supersymmetric and non-super\-sym\-metric vacuum solutions of the
$R^2$-extended Salam--Sezgin model discussed in the previous section. In
the first subsection we present the bosonic field equations of this
model. In the following three subsections we investigate vacuum
solutions with no fluxes, 2-form fluxes and 3-form fluxes, respectively.
In the last subsection
we compute the spectrum of the theory around six dimensional Minkowski
spacetime.


\subsection{Bosonic Field Equations}


For the purpose of finding the vacuum solutions, it is convenient to
eliminate the auxiliary fields as much as possible. Prior to adding the
Riemann tensor squared invariant, we saw that the auxiliary fields
$(E_{\mu\nu\rho\sigma}, {\cal V}_\mu^{\prime ij}, {\cal V}_\mu,Y^{ij})$ can
all be eliminated by using their field equations.  However, upon the
addition of the Riemann tensor squared invariant, while we can still eliminate
$(Y^{ij}, E_{\mu\nu\rho\sigma})$, we can no longer eliminate $({\cal
V}_\mu^{\prime ij}, {\cal V}_\mu)$ since they acquire kinetic terms. Thus, we
shall proceed with the elimination of $(Y^{ij}, E_{\mu\nu\rho\sigma})$
only. The relation
\be
Y^{ij}= -\frac{1}{2\sqrt 2} gL\delta^{ij}\ ,
\label{Yeq}
\ee
readily follows from \eq{Lag2}, while the $E_{\mu\nu\rho\sigma}$ field equation gives
\be
\varepsilon^{\lambda\tau\rho\sigma\mu\nu} \partial_\mu\left( L^{-1}E_\nu
-{\sqrt 2} {\cal V}_\nu - g W_\nu\right)=0 \ .
\label{Ceq}
\ee
This implies that we can locally write
\be
L^{-1}E_\mu -{\sqrt 2} {\cal V}_\mu - g W_\mu=\partial_\mu\phi\ ,
\ee
for some scalar $\phi$, which inherits the shift gauge symmetry transformations \eq{gt}. This
symmetry is readily fixed by setting $\phi$ equal to a constant, thereby arriving at the
field equation
\be
E_\mu = {\sqrt 2} L \left( {\cal V}_\mu + \frac{1}{\sqrt 2} g
W_\mu\right)\ .
\label{defE}
\ee
Taking into account \eq{Yeq} and \eq{defE}, we find the following bosonic field equations
for the propagating fields in the theory (\ref{totalbos}):
%
%
\bea
L R_{\mu\nu}&=& \nabla_\mu\nabla_\nu L - L^{-1}\partial_\mu L\partial_\nu L
+\frac14 g^2 g_{\mu\nu}L^2 +\frac14 L F_{\mu\rho\sigma}(B) F_{\nu}{}^{\rho\sigma}(B)
\nn\\
&& -4L Z_{(\mu} Z^*_{\nu)} -\frac12 L^{-1}E_\mu E_\nu
 +F_{\mu\rho}(W)F_\nu{}^{\rho}(W)
 \nn\w2
&&-\frac14 g_{\mu\nu} F_{\rho\sigma}(W)F^{\rho\sigma}(W)-\frac{1}{8M^2} S_{\mu\nu}\,, \label{Einstein}\w2
%
%
R &=& g^2 L +2 L^{-1} \Box L -L^{-2}\partial_\mu L \partial^\mu L
+\frac{1}{12} F_{\mu\nu\rho}(B)F^{\mu\nu\rho}(B) \nn\w2 && -4Z_\mu Z^{*\mu}
-\frac12 L^{-2} E_\mu E^\mu \,, \label{Leq}\w2
%
%
\nabla_\rho \left(L F^{\rho\mu\nu}(B)\right)  &=&
\frac14e^{-1}\varepsilon^{\mu\nu\rho\sigma\lambda\tau} \left(
F_{\rho\sigma}(W)F_{\lambda\tau}(W) +\frac{1}{2M^2} \tR^{\alpha\beta}{}_{\rho\sigma}
\tR_{\alpha\beta\lambda\tau}\right) \nn\w2 && +\frac{3}{M^2}
\nabla_\alpha\tn_\beta \tR^{[\mu\nu\alpha]\beta} +\frac{3}{M^2}
\nabla_\alpha \left(F^{-\,\rho\sigma[\alpha}(B)
\tR^{\mu\nu]}{}_{\rho\sigma}\right)\,, \label{Beq}\w2
%
%
0&=& \nabla_\mu F^{\mu\nu}(W) +\frac12 g E^\nu
+\frac12  \tilde F^{\nu\rho\sigma}(B) F_{\rho\sigma}(W)\ ,
\label{Aeq}\w2
%
%
0&=& \nabla_\nu F^{\mu\nu}({\cal V})+\left[2\rmi F^{\mu\nu}(Z) Z^*_\nu +h.c.\right] + \frac{1}{\sqrt 2} M^2  E^\mu\ ,
\label{Veq}\w2
%
%
0&=& (\partial_\mu -\rmi{\cal V}_\mu )F^{\mu\nu}(Z) -\rmi F^{\nu\rho}({\cal V}) Z_\rho -M^2LZ^\nu \ ,
\label{V'eq}
\eea
where $E^\mu$ is the $\U(1)$ invariant vector field determined in terms of
the vector fields $W_\mu$ and ${\cal V}_\mu$ as in \eq{defE}. The fact that $E^\mu$ is
divergence free follows from \eq{Aeq}, and separately from \eq{Veq}. We have
also defined
\bea
S_{\mu\nu} &\equiv& 8 F_{\mu\rho}({\cal V}) F_\nu{}^{\rho}({\cal V})
-2 g_{\mu\nu} F_{\rho\sigma}({\cal V})F^{\rho\sigma}({\cal V})
-32F_{\rho(\mu}(Z) {F^*}_{\nu)}{}^{\rho}(Z)
- 8 g_{\mu\nu} F_{\rho\sigma}(Z) F^{*\rho\sigma}(Z)
\nn\\
&& -4 \tR^{\lambda\tau}{}_{\mu\rho} \tR_{\lambda\tau\nu}{}^\rho
 +g_{\mu\nu} \tR_{\lambda\tau\rho\sigma}\tR^{\lambda\tau\rho\sigma}
+8\nabla^\alpha\tn^\beta \tR_{\alpha(\mu\nu)\beta}
+8\nabla^\alpha \left(\tR_{\alpha(\mu}{}^{\rho\sigma} F^-_{\nu)\rho\sigma}(B)\right)
\nn\\
&&
+4 F^\alpha{}_{\lambda(\mu}(B)  \tn^\beta \tR^\lambda{}_{\nu)\alpha\beta}
-4 \tR_{\lambda(\mu}{}^{\alpha\beta} F_{\nu)}{}^{\lambda\tau}(B)F^-_{\tau\alpha\beta}(B)\,,
\label{defS}
\eea
where $F^\pm(B) = (F(B) \pm \tilde F(B))/2$ with $\tilde F^{\mu\nu\rho}= -\ft16 e^{-1}\varepsilon^{\mu\nu\rho\sigma\lambda\tau}
F_{\sigma\lambda\tau}$. We
have simplified the Einstein equation by using \eq{defE} and the $L$ field equation \eq{Leq}. We have also used the definitions
\be \tR^\alpha{}_{\beta\mu\nu} = \del_\mu \widetilde
\Gamma^\alpha_{\nu\beta}+\cdots\,, \qquad
{\widetilde\Gamma}^\rho{}_{\mu\nu}\equiv \Gamma^\rho{}_{+\mu\nu } =
 \Gamma^\rho{}_{\mu\nu} + \ft12 F^\rho{}_{\mu\nu}(B)\ .
\ee
Thus, we have
\be
\widetilde R^{\alpha\beta}{}_{\mu\nu}= R^{\alpha\beta}{}_{\mu\nu} -
\nabla_{[\mu} F_{\nu]}{}^{\alpha\beta}(B) - \ft12 F^\alpha{}_{\lambda[\mu}(B)
F^{\beta\lambda}{}_{\nu]}(B)\ .
\label{tR}
\ee
Given the vielbein postulate
\begin{equation}
  \partial_\mu e_\nu^a
+\omega_{\mu\pm}{}^{ab} e_{\nu b} -\Gamma_\mp^\rho{}_{\mu\nu}\, e_\rho^a
= 0
 \end{equation}
with $\omega_{\mu\pm}{}^{ab}$ and $\Gamma^\rho{}_{\pm\mu\nu}$ defined in (\ref{spinconnbostors}) and (\ref{chrsymbbostors}),
respectively, it follows that
\be R_{\mu\nu}{}^{ab} (\omega_-) e^\lambda_a e_{\tau b} =
R^\lambda{}_{\tau\mu\nu} (\Gamma_+) \equiv \tR^\lambda{}_{\tau\mu\nu}\,.
\ee
 The occurrence of covariant derivatives with and without bosonic torsion in
the quantity $S_{\mu\nu}$ is due to the following manipulation:
\bea && \delta\int e R_{\mu\nu ab}(\omega_- ) R^{\mu\nu ab}(\omega_- )=
4\int R^{\mu\nu}{}_{ab}(\omega_- ) D_\mu(\omega_-)\delta\omega_{\nu
-}{}^{ab} + {\rm a\ term} \sim \delta(e g^{\mu\rho}g^{\nu\sigma})
\nn\\
&&= 4\int R^{\mu\nu}{}_{ab}(\omega_- )
\left[D_\mu(\omega_-,\Gamma_+)\delta\omega_{\nu -}{}^{ab}
 +\frac12 F_{\mu\nu}{}^\rho(B) \delta\omega_{\rho -}{}^{ab}\right] + {\rm a\ term} \sim \delta(e g^{\mu\rho}g^{\nu\sigma})\,.\nonumber\\
 &&
\eea
A partial integration in the first term is then responsible for the
occurrence of $\tn$ in the expression for $S_{\mu\nu}$. Another useful
variational formula takes the form
\bea
&& \delta \int \varepsilon^{\mu\nu\rho\sigma\lambda\tau} B_{\mu\nu}
R_{\rho\sigma}{}^{ab}(\omega_-) R_{\lambda\tau ab}(\omega_-) \w2 && =
\varepsilon^{\mu\nu\rho\sigma\lambda\tau} \left( \int (\delta
B_{\mu\nu}) R_{\rho\sigma}{}^{ab}(\omega_-) R_{\lambda\tau ab}(\omega_-)
+4 B_{\mu\nu} \partial_\rho \left[R_{\lambda\tau
ab}(\omega_-)\delta\omega_{\sigma -}{}^{ab}\right]\right)\ .\nn
\eea

The field equations for the abelian vector fields $W_\mu$ and ${\cal V}_\mu$
have an intricate structure. Suitable combinations of these fields describe
a gauge field $X_\mu $ and a gauge invariant Proca field $Y_\mu$ given by
\be
X_\mu \equiv  {\cal V}_\mu +{\sqrt 2} {g^{-1}M^2}\, W_\mu\ ,
\qquad Y_\mu
\equiv  {\cal V}_\mu + \frac{g}{\sqrt 2}\, W_\mu  \ . \ee
The field equations \eq{Aeq} and \eq{Veq} can then be written as
\bea
&& \nabla_\mu X^{\mu\nu} = \frac{M^2}{g^2-2M^2}\,
\tilde F^{\nu\rho\sigma}(B)( X_{\rho\sigma}- Y_{\rho\sigma} )\ ,
\label{Weq}\w2
&& \nabla_\mu Y^{\mu\nu} +\frac12 (g^2-2M^2)LY^\nu = \frac{g^2}{2(g^2-2M^2)}\,
\tilde F^{\nu\rho\sigma}(B)( X_{\rho\sigma}- Y_{\rho\sigma} )\ ,
\label{Zeq}
\eea
for $2M^2-g^2 \ne 0$, and $X_{\mu\nu}$, $Y_{\mu\nu}$ given by
\be
X_{\mu\nu}= \partial_\mu X_\nu -\partial_\nu X_\mu\ ,\qquad
Y_{\mu\nu}= \partial_\mu Y_\nu -\partial_\nu Y_\mu\ .
\ee
In the special case that $M^2=g^2/2$, the left hand side of the field equations \eq{Weq} and \eq{Zeq}
can no longer be diagonalized. As we saw earlier, this is a critical point at which the coefficient of the  kinetic term for the Maxwell vector field vanishes to lowest order in $1/M^2$ when the auxiliary vector field ${\cal V}_\mu$ is eliminated to the same order.


\subsection{Vacuum Solutions Without Fluxes}


If $g\ne 0$, the field equations do not admit a single constant curvature $6D$
spacetime solution for any value of the constant curvature, with or
without
supersymmetry. In particular, Minkowski spacetime is not a solution as can be
readily seen from the equation $R=g^2 L_0$, where $L=L_0$ is a non-vanishing
constant and all other fields are set equal to zero. If $g^2=0$, on the other
hand, setting $L$ equal to a constant and all the other fields equal to
zero yields Minkowski$_6$ as a supersymmetric solution.

Next, let the six dimensional spacetime be a direct product of constant
curvature spaces $M_1\times M_2$, with dimensions $d_1$ and $d_2$. We find that solutions exist with
\bea && R_{\mu\nu\rho\sigma} =  \frac{n_1}{d_1(d_1-1)} g^2 L_0\,
(g_{\mu\rho}g_{\nu\sigma}-g_{\mu\sigma}g_{\nu\rho})\ ,\ \ R_{pqrs} = \frac{n_2}{d_2(d_2-1)} g^2
L_0( g_{pr}g_{qs}-g_{ps}g_{qr})\ , \nn\w2 && L = L_0\ ,\quad M^2 = \frac{1}{2}n_3 g^2\ ,
\label{curvatures1} \eea
with all the other fields vanishing. Here $L_0$ is an arbitrary non-vanishing
positive constant, and the numbers $(n_1,n_2,n_3)$ are given in Table \ref{m1m2}.
Note that here we are using the coordinates $(x^\mu,y^r)$.

\begin{center}
\begin{table}[ht]
\caption{Solutions of the form $M_1\times M_2$ in the absence of fluxes. The numbers $(n_1,n_2,n_3)$
are defined in
(\ref{curvatures1}).\\}
\label{m1m2}
\centering
\begin{tabular}{|c|c|c|c|}
\hline
Spacetime & $n_1$ & $n_2$ & $n_3$ \\[.2truecm]
\hline\rule[-1mm]{0mm}{6mm}
Mink$_4\times S^2$ & 0 & 1 & 1 \\[.2truecm]
\rule[-1mm]{0mm}{6mm}
dS$_4\times T^2$ &  1 & 0 & 1/6 \\[.2truecm]
\rule[-1mm]{0mm}{6mm}
dS$_4\times S^2$ &  6/7 & 1/7 & 1/7 \\[.2truecm]
\hline\rule[-1mm]{0mm}{6mm}
Mink$_3\times S^3$ & 0 & 1 & 1/3 \\[.2truecm]
\rule[-1mm]{0mm}{6mm}
dS$_3\times T^3$ &  1 & 0 & 1/3 \\[.2truecm]
dS$_3\times S^3$ &  1/2 & 1/2 & 1/6\\[.2truecm]
\hline
\end{tabular}
\end{table}
\end{center}

There are also solutions involving a product of three 2-dimensional constant
curvature spaces, whose curvature constants, allowed to vanish as well, are
chosen properly. In all these solutions, and those tabulated above, $M^2$ is
fixed in terms of $g^2$, and all solutions are non-supersymmetric.


\subsection{Vacuum Solutions With 2-Form Flux}


Next, let us consider a spacetime $M_4 \times M_2$, which is a direct
product of two constant curvature spaces and turn on the fluxes produced
by $F(W)$ and $F({\cal V})$ on $M_2$. We set $L$ equal to a positive
non-vanishing constant and the remaining fields equal to zero. In
particular, from \eq{defE} it follows that ${\cal
V}_\mu=-gW_\mu/\sqrt{2}$. Using this information, we can make the
following Ansatz for the non-vanishing fields:
\bea
&& R_{\mu\nu}= 3 a\, g_{\mu\nu}\ ,\qquad R_{rs} = b\, g_{rs}\ ,
\qquad L=L_0\ , \nn\w2
&& F_{rs}(W)= c\, \sqrt{g_2}\,\varepsilon_{rs}\ , \qquad
F_{rs}({\cal V})=-\frac{g}{\sqrt 2}\, c \sqrt{g_2}\,\varepsilon_{rs}\ ,
\label{sss}
\eea
where $a,b,c, L_0$ are constants, $g_2 = \det g_{rs}$, we have used the coordinates $(x^\mu,
y^r)$ and
$\varepsilon_{12}=\varepsilon^{12}=1$. Using this ansatz we find the following solutions. One of them is a direct
product of $4D$ Minkowski spacetime with a 2-sphere, given by
\bea {\rm Mink}_4 \times S^2\ :\ \qquad a=0\ , \qquad b=\frac12 g^2 L_0\ ,\qquad c=\pm \frac{g L_0}{\sqrt 2}\ .
\label{solution2}
\eea
Remarkably, this is precisely the supersymmetric Salam--Sezgin solution
for any value of $M^2!$
For this solution, the integrability condition for the Killing spinor equation $\delta_\epsilon\psi_{\hat \mu}=0$ is
\begin{equation}
 \left[ R_{\hat\mu\hat\nu \hat a\hat b}\Gamma^{\hat a\hat b}\varepsilon^{ij}  -2 F_{\hat\mu\hat\nu}({\cal V})\delta^{ij}\right]\epsilon_j=0 \ ,
\end{equation}
where $\hat\mu, \hat a=0,1,...,5$. For the solution \eq{solution2} this gives\footnote{We
decompose \label{fn:D64spinors}
the $6D$ Dirac matrices as $\Gamma_\mu=\gamma_\mu \otimes 1,
\Gamma_4=\gamma_* \otimes \sigma_1$ and
$\Gamma_5=\gamma_*\otimes \sigma_2$. Then $\Gamma_*=\gamma_*
\otimes \sigma_3$.
 This defines 4-dimensional spinors $\epsilon _{Ai}=\gamma _*(\sigma _3)_A{}^B\epsilon _{Bi}$,
where the 4-dimensional spinor index is suppressed and $A,B=1,2$ labels the
2-dimensional spinors on $S^2$. The combinations
$\eta _1= \epsilon _{11}+\rmi\epsilon _{22}$ and $\eta _2=\epsilon _{12}-\rmi\epsilon_{21}$
are 4-dimensional Majorana spinors.}
\be
\rmi\left(\sigma_3\right)_A{}^B\,\delta _{ik}\varepsilon^{kj} \,\epsilon_{Bj} =
\mp \epsilon_{Ai}\,.
\label{susyc}
\ee
%
The vanishing of $\delta _\epsilon \varphi^i$ follows trivially, and, using
(\ref{solution2}) and (\ref{susyc}), it follows that  $\delta_\epsilon  \Omega^i=0$ as
well. So the only independent condition on the Killing spinor is given by
(\ref{susyc}). It implies $\mathcal{N}=1$
supersymmetry in Minkowski$_4$. Indeed, using the Majorana spinors $\eta _1$
and $\eta _2$ defined in footnote \ref{fn:D64spinors},
%
%
the condition \eq{susyc} turns into $\rmi\gamma_*\eta_1=\pm \eta_2$.

The other solutions are given by
\bea
a &=&M^2L_0\ ,\qquad b=\frac12 (g^2-12M^2) L_0\ , \nn\w2
 c &=& \pm
\,L_0\, \sqrt{\frac{(g^2-12M^2)(g^2-14M^2)}{2(g^2-2M^2)}}  \ ,\qquad
M^2\ne \frac12 g^2\ ,
\label{f2}
\eea
and therefore they describe the following spaces:
\bea && AdS_4 \times S^2\ :\ \qquad M^2 <0\ , \label{sol1}\w2 && dS_4 \times S^2\ :\
\qquad \frac{1}{14} g^2 >M^2 >0\ ,
\label{sol2}\w2
&& dS_4 \times H^2\ :\ \qquad   \frac12 g^2 >  M^2 > \frac1{12} g^2\ ,
\label{sol3}
\eea
where $H^2$ is a 2-hyperboloid. For the special case of $M^2=g^2/2$, there exists the following solution
\be
{\rm Mink}_4 \times S^2\ :\ \qquad a=0\ , \qquad b=\frac12 g^2 L_0\ ,\qquad   M^2=\frac12
g^2\ ,
\ee
for \emph{any value of $c$}, which contains as a special case the solution \eq{solution2} for $c=\pm gL_0/{\sqrt 2}$, and the first entry in Table \ref{m1m2} for $c=0$. Of all the solutions with the 2-flux turned on, the only supersymmetric one is the one given in \eq{solution2}.


\subsection{Vacuum Solutions With 3-Form Flux}


We shall take the $6D$ spacetime to be a direct product of two three-dimensional constant
curvature spaces $M_1\times M_2$ with coordinates $(x^\mu,y^r)$, set
$L=L_0$,
turn on the 3-form flux and set the remaining fields equal to zero. Thus we
have
\bea R_{\mu\nu}{}^{\rho\sigma} &=& 2a\, \delta_{[\mu}^\rho
\delta_{\nu]}^\sigma ,\qquad R_{pq}{}^{rs} = 2b\,\delta_{[p}^r\delta_{q]}^s
,\qquad L=L_0\ , \nn\w2 F_{\mu\nu\rho}(B) &=& 2c_1 \sqrt{-g_1}\varepsilon_{\mu\nu\rho}\ ,
\qquad F_{rst}(B) = 2 c_2 \sqrt{g_2} \varepsilon_{rst}\ ,
 \label{an33}
 \eea
where $g_1=\det g_{\mu\nu}$ and $g_2=\det g_{rs}$. From \eq{tR} we get
\be \tR_{\mu\nu}{}^{\rho\sigma} = 2(a +c_1^2 )\,\delta_{[\mu}^\rho
\delta_{\nu]}^\sigma\ , \qquad \tR_{pq}{}^{rs} = 2(b-c_2^2)\,
\delta_{[p}^r\delta_{q]}^s\ . \label{a33} \ee
If we set $g^2=0$, then all the terms that depend on $M^2$ vanish since the
curvatures defined above vanish due to the non-vanishing (parallelizing)
torsion. This gives the known $AdS_3\times S^3$ solution
\be
AdS_3\times S^3 \ : \ \qquad c_1^2=c_2^2=-a=b\ .
\label{3flux}
\ee
This solution preserves full supersymmetry. Indeed the integrability
condition for the existence of Killing spinors requires that the
torsionful curvatures vanish, and this is the case with the 3-form
fluxes as given in \eq{3flux}. As a consequence, all the contributions
of the Riemann tensor squared invariant to the field equations vanish in this case.

Next, we seek solutions with $g^2\ne 0$ and nonvanishing 3-form
flux. To bring the field equations to a manageable form, we shall
supplement the Ansatz \eq{an33} with a further condition and introduce
some notation
\be c_1=-c_2 \equiv c\ .\qquad \ee
Finding the most general such solution yields rather complicated relations
among the parameters. However, we have managed to find the following
relatively simple and intriguing solutions:
\be a= \frac{1}{6}(-6c^2+g^2L_0)\ ,\quad
                          b= c^2\ ,\quad M^2=\frac{g^2}{6}\,,
\ee
for arbitrary $c^2>0$. This solution corresponds to $dS_3\times S^3$ for $0<c^2<\frac{g^2L_0}{6}$ and to $AdS_3\times S^3$ for $c^2>\frac{g^2L_0}{6}$. Another solution is given by
\bea
a_{\pm}&=&\frac{1}{24}\Bigl(5g^2L_0-24L_0M^2\mp\sqrt{3}\sqrt{L_0^2(g^4-12g^2M^2+48M^4)}\Bigr)\,, \nonumber \\
b_{\pm}&=&\frac{1}{24}\Bigl(-g^2L_0+24L_0M^2\pm\sqrt{3}\sqrt{L_0^2(g^4-12g^2M^2+48M^4)}\Bigr)\,, \nonumber \\
c_{\pm}^2&=&\frac{1}{24}\Bigl(g^2L_0\mp\sqrt{3}\sqrt{L_0^2(g^4-12g^2M^2+48M^4)}\Bigr)\,,
\eea
where the $+$ solution corresponds to $dS_3\times S^3$ for $\frac{g^2}{12}<M^2<\frac{g^2}{6}$ and the $-$ solution corresponds to $AdS_3\times S^3$ for $M^2>\frac{11g^2}{36}$ and to $dS_3\times H^3$ for $M^2<\frac{g^2}{12}$.
These solutions are non-supersymmetric.


\subsection{Spectrum in Minkowski Spacetime}


Setting  $g^2=0$, and expanding around $6D$ Minkowski spacetime, we define
the linearized fluctuations
\begin{equation}
 g_{\mu\nu}=\eta_{\mu\nu}+h_{\mu\nu}\,,\qquad L=L_0+ \phi\ .
\end{equation}
Since all the other background fields are vanishing, we find that the
linearized Einstein and $L$ field equations take the form
\bea 0&=&  L_0 \left(\Box h_{\mu\nu} +\partial_\mu\partial_\nu h
-2\partial_{(\mu}\partial^\alpha h_{\nu)\alpha}\right)
+2\partial_\mu\partial_\nu \phi
\nn\\
&& -\frac{1}{M^2} \left(\Box\Box h_{\mu\nu} -2\Box
\partial_{(\mu}\partial^\alpha h_{\nu)\alpha}
+\partial_\mu\partial_\nu\partial^\alpha\partial^\beta
h_{\alpha\beta}\right)\ ,
\label{ein}\w2
 0 &=& L_0\left(\Box
h-\partial^\mu\partial^\nu h_{\mu\nu}\right) + 2\Box \phi\ . \label{sc} \eea
Note that we have not imposed any gauge conditions yet. Using \eq{sc} in the
trace of \eq{ein}, we find
\be
\Box\left(\Box -M^2L_0 \right)\phi = 0\ .
\label{phe}
\ee
To simplify Einstein's equation, however, it is
convenient to impose the gauge condition
\be
\partial^\mu h_{\mu\nu}=\frac12 \partial_\nu h \ .
\label{gc} \ee
In this gauge, the trace of Einstein's equation and \eq{sc} give
\bea
\Box\left(\Box - M^2L_0 \right) h &=& 0\ ,
\label{he}\w2
 \Box h &=& -4L_0^{-1}\Box \phi\ .
\label{ce}
\eea
We shall assume that $M^2\ne 0$. Then it follows from \eq{phe} that either $(\Box-M^2 L_0)\phi=0$ or $\Box\phi=0$. In the first case, defining $\chi \equiv \Box\phi $, it follows from \eq{phe}, \eq{he} and \eq{ce} that there is one massive scalar obeying $(\Box -M^2 L_0)\chi=0$. In the latter case, $\Box\phi=0$ and it follows from \eq{ce} that $\Box h=0$ as well. However, the solution of $\Box h=0$ can be gauged away by the residual general coordinate transformations that preserve the gauge condition \eq{gc}. Thus, we are left with a massless scalar described by $\Box\phi=0$.

Turning to Einstein's equation, using the gauge condition \eq{gc}, and the field equations obeyed by $h$ and $\phi$, it becomes
\be
\left(\Box-M^2 L_0 \right)\Box h_{\mu\nu}=
-2L_0^{-1}\left(\Box-M^2 L_0 \right)\partial_\mu\partial_\nu \phi\ .
\label{ein2}
\ee
This equation, when $(\Box-M^2 L_0)\phi=0$, reduces to $\left(\Box-M^2 L_0 \right)\Box h_{\mu\nu}=0$, describing a massless graviton and
 a massive graviton with mass $M\sqrt{L_0}$,
one of which, depending on the overall sign in the action, has the wrong sign kinetic
term.
 If $\Box\phi=0$, then we have
$\left(\Box-M^2 L_0 \right)\Box h_{\mu\nu}= -2M^2\partial_\mu\partial_\nu \phi$. In this case, the solution of  $\Box\phi=0$ is to be substituted to the right hand side of this equation and treated as a given external source. Note that the gravitational field does not appear as a source in the field equation for the scalar $\phi$, and there is no diagonalization problem here. Thus, the equation \eq{ein2} again describes a massless and ghost massive graviton of mass $M\sqrt{L_0}$.

The remaining field equations in the usual Lorentz gauges take the form
\be
  \Box a_\mu= 0\ , \qquad
   \left(\Box
-M^2L_0 \right)\left(
                                               \begin{array}{c}
                                                 v_\mu \\
                                                 z_\mu \\
                                               \end{array}
                                             \right)=0\ , \qquad
\Box (\Box -M^2 L_0) b_{\mu\nu}=0\,,
\ee
where the notation for the fluctuations is self explanatory. These
equations describe a massless vector and 2-form potential, a massive ghostly
2-form potential and three massive ghostly vectors.

Next, we examine the linearized fermion field equations. Imposing the gauge condition $\gamma^\mu\psi_\mu=0$, and
defining\footnote{This $\psi^i $ is unrelated to the $\psi^i$ introduced in (\ref{Weyl}), which was eliminated by (\ref{gf2}).}
$\psi^i\equiv \partial^\mu\psi^i_\mu$, a straightforward manipulation of the fermion field equations gives
\bea
&& \slashed{\partial} (\Box-M^2L_0)\psi'_\mu = 0\ ,\qquad \slashed{\partial} \Omega=0\ ,
\label{ferm1}\w2
&&\Box\psi_i = {\sqrt 2} M^2 \slashed{\partial} \varphi^j \delta_{ij}\ ,
\qquad \psi_i = {\sqrt 2}L_0^{-1} \slashed{\partial} \varphi^j \delta_{ij}\ ,
\label{ferm2}
\eea
where $\psi '_\mu \equiv \psi _\mu -\Box^{-1} \partial _\mu \partial _\nu \psi
^\nu $, i.e.
 $\partial^\mu {\psi'}_\mu=0$. From \eq{ferm2}, it follows that $\slashed{\partial} \left(\Box-M^2 L_0 \right)\varphi=0$.
Therefore, altogether we have a massless gravitino, tensorino $\varphi$ and
gaugino together with a
massive gravitino and tensorino, both with mass $M\sqrt{L_0}$.

In summary, the
full spectrum consists of the massless Maxwell multiplet with fields $(a_\mu,\, \Omega)$, the
(reducible) massless $16+16$ supergravity multiplet with fields
$(h_{\mu\nu},\, b_{\mu\nu}, \,\phi,\, \psi_\mu,\,\varphi)$ and a massive $40+40$ supergravity
multiplet of ghosts with fields $(h_{\mu\nu},\, b_{\mu\nu},\, z_\mu,\,
v_\mu,\,\phi,\,\psi_\mu,\,\varphi)$, all with the same mass, $M\sqrt{L_0}$, as expected.

\section{Conclusions} \label{section: Conclusions}
Our main goal in this paper has been the study of the R-symmetry gauging
in the presence of higher derivative corrections to Poincar{\'e}
supergravity and its consequences for the vacuum solutions. To this end,
we first studied the gauging of the U$(1)$ R-symmetry of $\mathcal{N} =
(1, 0)$, $D = 6$ supergravity in the off-shell formulation. The
off-shell Poincar{\'e} supergravity theory already has a local $\U(1)_R$
symmetry but it is gauged by an auxiliary vector field which is not
dynamical. We performed the gauging that employs a dynamical gauge field
by coupling the model to an off-shell vector multiplet equipped with its
own $\U(1)$ symmetry. Then, we showed that this model has a shift
symmetry which can be fixed, thereby breaking $\U(1)_R \times \U(1)$
down to a diagonal $\U(1)_R^{\rm diag}$. As a result the auxiliary vector gets
related to the vector coming from the Maxwell multiplet, and the
on-shell model obtained in this manner agrees with the dual formulation
\cite{Nishino:1984gk} of the gauged Einstein--Maxwell supergravity
constructed long ago \cite{Salam:1984cj,Nishino:1984gk}.

Next, we added a curvature squared supersymmetric invariant, with
the Riemann tensor squared as its leading term, to the off-shell model and studied its influence on the gauging procedure. This invariant
causes the auxiliary fields to become `propagating' and to mix with the
physical fields. A particular combination of the physical vector
and the auxiliary vector gauges the symmetry and another combination describes a massive vector field inert under $\U(1)$. We can, however, put a small
parameter in front of the curvature squared part of the Lagrangian and
consider it as a higher-order correction term. Then the auxiliary fields
can be eliminated order by order and the gauging proceeds again via the
vector field residing in the Maxwell multiplet. Treating  the higher derivative extension either way, we have seen that the positive definite potential that arises in the minimal model does not get modified.

Chiral gauged supergravity in six dimensions is known to admit a
(supersymmetric) chiral ${\rm Minkowski}_4\times S^2$ compactification,
while it does not admit a six-dimensional Minkowski or (anti) de Sitter
spacetime as a solution, regardless of supersymmetry
\cite{Salam:1984cj}. We have shown that the inclusion of the Riemann
tensor squared invariant remarkably leaves the supersymmetric  ${\rm
Minkowski}_4\times S^2$ solution intact. We have also found new solutions
in which the spacetime and the internal spaces may have positive or
negative curvature constants. It is noteworthy that de Sitter spacetime
solutions exist, avoiding a no go theorem that exists for ten
dimensional  supergravities\footnote{Note that a possible string theory embedding does not contradict the avoidance of the 10$D$ no go theorem since this theorem no longer holds when higher derivative corrections are included.} \cite{Gibbons:1984kp,Maldacena:2000mw}.
While the spectrum in the 2-sphere compactification remains to be
determined, we have found that the spectrum of the ungauged theory in
six dimensional Minkowski spacetime, not surprisingly, has a ghostly
massive spin two multiplet in addition to a massless supergravity and a
Maxwell vector multiplet.

Given that the $(1,0)$ supergravity theory in six dimensions is the most
supersymmetric and highest-dimensional supergravity model that admits an
off-shell formulation, and that it admits an exactly supersymmetric
higher derivative extension, it is worthwhile to study this model
further. The coupling of Yang-Mills and hypermultiplets would be useful.
In particular, a possible modification of the quaternionic K{\"a}hler
geometry, and consequences for the compactification would be interesting
to determine. The model without such couplings harbors many anomalies.
It is important to study the gravitational, gauge and mixed anomalies in
the matter-coupled version of the higher derivative extended theory. The
Green-Schwarz  anomaly counterterm that involves the gravitational
Chern--Simons term arises as part of the Riemann tensor squared
invariant. However, the presence of the Riemann tensor squared term
raises the question with regard to the presence of ghosts in the
spectrum, defined in the presence of a suitable vacuum solution. Indeed,
dealing with the ghost problem is of great importance for this model to
have applications to model building, and it remains to be investigated.
In particular, the consequences of the higher derivative extension for
the braneworld scenarios put forward in \cite{Aghababaie:2003wz} where
3-branes are inserted at singular points of the 2-dimensional internal
space, would be worthwhile to explore.

Various properties of the model we have studied here would naturally be
affected by the presence of an additional higher-derivative
supersymmetric invariant. In five dimensions, for example, it is known
that a Weyl tensor squared invariant exists, in addition to the Riemann
tensor squared invariant, which can be obtained from a circle reduction
of the one studied here. However, whether the Weyl tensor squared or
another combination of the curvature squared terms can be
supersymmetrized in six dimensions is an open problem. If such
invariants exist, not only would they be useful in avoiding the ghost
problem, they would also play a significant role in a possible embedding
of these theories, albeit in the ungauged setting, to the string theory
low energy effective action. For a preliminary discussion of this problem,
in the context of the Riemann tensor squared model we already have, see
\cite{Lu:2010ct}.

The embedding of the higher-derivative extended model to string theory
might also provide new grounds for testing the conjectured connection
between microscopic and macroscopic black hole entropy. The use of
off-shell supersymmetric Riemann tensor squared extended $\mathcal{N}=2$, $D=4$
supergravity in this respect has been illustrated in
\cite{LopesCardoso:1998wt}. The existence of static, rotationally
symmetric black hole solutions that are $\mathcal{N}=2$ supersymmetric and that
approach Minkowski spacetime at spatial infinity and Bertotti-Robinson
spacetime at the horizon play a significant role in the work of
\cite{LopesCardoso:1998wt}. It is notoriously difficult to find exact
black hole solutions of higher-derivative gravities. Black hole
solutions of the ungauged $(1,0)$ $6D$ supergravity have been found in
\cite{Gibbons:1987ps}  and there exists an exact string solution of the theory we have studied in this
paper \cite{Lu:2010cg}. Nevertheless, black hole solutions in the presence of
gauging and/or a higher-derivative extension remains an open and
challenging problem.

\section*{Acknowledgments}
This work is supported in part by the FWO - Vlaanderen, Project No.
G.0651.11, and in part by the Federal Office for Scientific, Technical
and Cultural Affairs through the ``Interuniversity Attraction Poles
Programme -- Belgian Science Policy'' P6/11-P. We thank Hong Lu, Yi Pang
and Chris Pope for useful discussions. The research of E.~S. is
supported in part by NSF grant PHY-0906222.

\newpage
\providecommand{\href}[2]{#2}\begingroup\raggedright\endgroup

\end{document}